\documentclass[journal,10pt,twocolumn,final]{IEEEtran}

\IEEEoverridecommandlockouts

\usepackage{amsmath,graphicx}
\usepackage{mathtools}
\usepackage{amssymb}
\usepackage{amsthm}
\usepackage{graphicx}
\usepackage{color}
\usepackage{subcaption}
\usepackage{tikz}
\usepackage{float}
\usepackage[hidelinks]{hyperref}
\usepackage{optidef}
\usepackage{cite}

\newcommand{\hide}[1]{\ifthenelse{\boolean{false}}{#1}{}}

\newtheorem{theorem}{{\bf Theorem}}

\newtheorem{lemma}{{\bf Lemma}}

\newtheorem{definition}{Definition}

\DeclareMathOperator*{\argmin}{arg\,min}

\newcommand{\brac}[1]{\left({#1}\right)}

\DeclareMathOperator*{\argmax}{\arg\!\max}

\newcommand{\Prob}[1]{\text{p}\brac{#1}}
\makeatletter

\newcommand{\Rmnum}[1]{\expandafter\@slowromancap\romannumeral #1@}
\makeatother
\newcommand{\vecsym}[1]{\mathbf{#1}} 

\newcommand{\covyy}[0]{\boldsymbol{\Sigma}_{yy}}
\newcommand{\mm}[0]{\boldsymbol{\Phi}}

\definecolor{resp1text}{RGB}{1, 150, 32} 
\definecolor{resp2text}{RGB}{26, 10, 180}    
\definecolor{resp3text}{RGB}{255,0,0}

\usepackage{etoolbox}
\makeatletter
\expandafter\patchcmd\csname\string\IEEEbiography\endcsname{plus 1fil}{plus 0pt}{}{}%
\makeatother

\begin{document}

\title{Sparse Bayesian Learning Algorithms Revisited: From Learning Majorizers to Structured Algorithmic Learning using Neural Networks}
%
%
%

\author{Rushabha Balaji,
        Kuan-Lin Chen, Danijela Cabric, and Bhaskar D. Rao\thanks{Rushabha Balaji and Danijela Cabric are with the Department of Electrical and Computer Engineering, University of California, Los Angeles, CA 90095, USA (e-mail: rubalaji99@g.ucla.edu, danijela@ee.ucla.edu). Kuan-Lin Chen and Bhaskar D. Rao are with the Department of Electrical and Computer Engineering, University of California, San Diego, La Jolla, CA 92093, USA (e-mail:
 kuc029@ucsd.edu; brao@ucsd.edu). 
This work was supported by NSF award 2224322. Bhaskar D. Rao and Kuan-Lin Chen were supported by the U.S. National Science Foundation under Grant CCF-2225617.}
}

\maketitle

\begin{abstract}
Sparse Bayesian Learning (SBL) is one of the most popular sparse signal recovery methods, and various algorithms exist under the SBL paradigm. However, given a performance metric and a sparse recovery problem, it is difficult to know a-priori the best algorithm to choose. This difficulty is in part due to a lack of a unified framework to derive SBL algorithms. In this work, we address this issue by first showing that the most popular SBL algorithms can be derived using the majorization-minimization (MM) principle. Hence, providing convergence guarantees to this class of SBL methods hitherto unknown. Moreover, we show that the two most popular SBL update rules not only fall under the MM framework but are both valid descent steps for a common majorizer, revealing a deeper analytical compatibility between these algorithms. Using this insight and properties from MM theory we expand the class of SBL algorithms, and address the question of finding the best SBL algorithm via data within the MM framework. Second, we go beyond the MM framework by introducing the powerful modeling capabilities of deep learning to further expand the class of SBL algorithms, and aim to learn a superior SBL update rule from data. We propose a novel deep learning architecture that can outperform the classical MM based ones across different sparse recovery problems. Our architecture is designed such that its complexity does not scale with the dimension of the measurement matrix, and hence provides us with a unique opportunity to test the generalization capability across different measurement matrices.  In the case of parameterized dictionaries, the invariance to the size of the matrix allows us to train and test the model across different ranges of the parameter. We also showcase our models ability to learn a functional mapping by its zero-shot performance on unseen measurement matrices. Finally, we test our model performance across different number of snapshots, signal-to-noise ratios and sparsity levels. 
\end{abstract}

%
\IEEEpeerreviewmaketitle

\section{Introduction}
We consider the  sparse signal recovery (SSR) problem in the multiple measurement vector (MMV) setting. It consists of a set of $L$ measurement vectors \cite{Cotter:MMV}, where the $l^{\text{th}}$ observation $\mathbf{y}_l\in \mathbb{R}^N$ is given by
\begin{equation}
    \label{eq:mm_model}
    \mathbf{y}_l = \boldsymbol{\Phi}\mathbf{x}_l + \mathbf{n}_l,\; \forall\; l\in \{1,2,\ldots,L\}.
\end{equation}
Here $\boldsymbol{\Phi}\in \mathbb{R}^{N\times M}$ is the measurement matrix with $N<<M, L<N$, and $\mathbf{x}_l \in \mathbb{R}^M$ is an independent and identically distributed (i.i.d.) $k$-sparse vector with common support across the measurements. The noise vectors $\mathbf{n}_l \in \mathbb{R}^N$ are i.i.d. and sampled from a standard Gaussian distribution~$\mathcal{N}(0,\sigma^2\mathbf{I}_{N})$. $\mathbf{x}_l$ and $\mathbf{n}_l$ are independent of each other. This problem has been an important research topic with many applications, e.g., biomagnetic imaging \cite{owen2008estimating}, direction of arrival (DoA)~\cite{cs_survey} and wireless channel estimation~\cite{cs_mmwave}, among others. Concisely, let us denote the set of measurements as $\boldsymbol{Y}=[\mathbf{y}_1,\ldots,\mathbf{y}_L]$ and the set of sparse vectors as $\boldsymbol{X}=[\mathbf{x}_1,\ldots,\mathbf{x}_L]$. The SSR problem is an inverse problem where the solution to~\eqref{eq:mm_model} is given by solving the NP-hard problem 
\begin{mini}|s|
{\boldsymbol{X}}{||\boldsymbol{X}||_{0,1}}
{}{}
\label{eq:inverse_problem}
\addConstraint{||\boldsymbol{Y}-\boldsymbol{\Phi}\boldsymbol{X}||_{F}\leq \eta.}{}
\end{mini}
The $\ell_0$ norm on the rows of a matrix $\boldsymbol{X}$ is denoted by $||\cdot||_{0,1}$ which represents the number of non-zero rows in $\boldsymbol{X}$. To solve the SSR problem tractably, different methodologies have proposed different types of relaxations. 
Many popular algorithms, rooted in classical optimization theory and statistics, have been proposed to solve the SSR problem. Recently, with the advent of deep learning, data driven methods have also been utilized to solve it as well. We give a brief exposition of the two very different methodologies and highlight the unique advantages provided by each.

\textit{Traditional SSR approaches:} Imposing the $\ell$-$1$ regularization leads to the celebrated least absolute shrinkage and selection operator (LASSO) formulation of the SSR problem. Different iterative techniques have been proposed~\cite{omp,iht,Irina:97,Chartrand,Candes:reweightedl1,wipf} to solve these formulations of which the matching pursuit (MP) and  LASSO approaches have been heavily studied and characterized~\cite{foucart2013mathematical}. From a Bayesian perspective these methods fall under the type-I framework where sparsity inducing priors are imposed on $\boldsymbol{X}$ as suitable regularizers~\cite{wipf2011latent}. Sparse Bayesian learning (SBL) introduces the type-II Bayesian framework~\cite{wipf,tipping,giri2016type} and re-formulates the SSR problem as one of solving the non-convex maximum likelihood problem in the hyper-parameter domain. SBL implicitly regularizes the solution to be sparse and various iterative algorithms have been proposed to solve the non-convex objective function of SBL~\cite{tipping,wipf,wipf2011latent,wipf2007new,wipf2010iterative,peter,nannuru2019sparse}.

Every traditional SSR algorithm involves iteratively refining an estimate of $\boldsymbol{X}$ say $\boldsymbol{\hat{X}}$ until a stopping criterion is met. A fixed mapping is used as one moves from $\boldsymbol{\hat{X}}_j$ an estimate at the $j^{\text{th}}$ iteration to the next iteration, that is, the mapping $h$ such that $h(\boldsymbol{\Phi},\boldsymbol{Y},\boldsymbol{\hat{X}_j})=\boldsymbol{\hat{X}}_{j+1}$ is fixed across iterations.  Many techniques derived from different principles have  achieved this mapping including popular SBL algorithms, which forms the basis of this paper.

\textit{ Deep Learning for SSR: }
With the advent of deep learning (DL), complex relationships can be modeled by training a neural network architecture with the correct inductive bias. This feature is attractive for SBL algorithm development because the underlying objective function is non-convex and no optimization algorithms are known that guarantee a global minimum with reasonable complexity. End-to-end deep learning approaches, that is, learning a mapping $f_{\boldsymbol{\Phi}}$ such that $f_{\boldsymbol{\Phi}}(\boldsymbol{Y})\approx\boldsymbol{X}$ for a fixed measurement matrix, have been proposed in~\cite{wipf_fcnn,wipf_lstm} which warrants re-training if the measurement matrix changes. A structured extension of the DL framework to the SSR problem is to learn a mapping $h_{\theta}$ such that $h_{\theta}(\boldsymbol{\Phi},\boldsymbol{Y},\boldsymbol{\hat{X}}_j)=\boldsymbol{\hat{X}}_{j+1}$ and $\boldsymbol{\hat{X}}_{j+1}\to\boldsymbol{X}$ as $j\to J$ for some $J\in\mathbb{N}$, where $\theta$ is the set of learnable parameters. Prior works have incorporated DL to improve the SBL algorithm by using the deep unrolling paradigm~\cite{deep_unroll}. For example, the expectation maximization (EM)-based SBL algorithm was unrolled with spatial correlation for mmWave channel estimation in~\cite{qualcomm}. The focus of the paper is different in that the emphasis is on the application-specific considerations and not the question of improving a general SBL algorithm. In~\cite{chandra}, the authors address this question but the complexity of the architecture they propose increases with the number of snapshots, and the size of the measurement matrix. It warrants training a different model every time the number of snapshots change or if the size of the measurement matrix changes. For a fixed $L$ and $M$, the authors showcase the generalization capabilities of the model to matrices whose elements are sampled from the standard Gaussian distribution but performance of the model beyond this class of matrices is unknown.

In this work, we propose a neural network architecture designed to learn an SBL algorithm, positioning it as a competitive candidate among popular methods for solving the SBL objective function \footnote{This work is an extension of our work~\cite{rbalaji_icassp} which was presented at ICASSP 2025.}. Introducing the DL capabilities to SBL is a natural progression, where DL acts as a meta-learning framework, allowing SBL to improve its learning methodology and adapting it to the priors in the dataset. The two most popular SBL algorithms are the expectation-maximization (EM) based SBL algorithm~\cite{wipf} and Tipping's multiplicative update (MU) rule proposed in the original SBL work~\cite{tipping} and referred to herein as MU-SBL. As the network is intended to learn an algorithm, we optimize over the space of SBL algorithms, rather than learning a direct mapping from $\boldsymbol{Y}$ to $\boldsymbol{X}$. Similar to how the EM-SBL algorithm and MU-SBL algorithm are agnostic to the type of the sensing matrix, the size of the sensing matrix, the signal-to-noise ratio (SNR) of the measurements, and the number of measurements, we posit that for a neural network to be a competitive candidate as an SBL algorithm it should have the same properties as that of the classical ones. Hence, the inductive bias of such a network should ensure that the model learns a mapping that generalizes to unseen measurements and preferably to unseen measurement matrices.

To design such a network, we need to align the inductive bias of the network with a fundamental property that characterizes all classical SBL algorithms. To understand and formulate such a property is difficult since the two most popular SBL algorithms, the EM-SBL and the MU-SBL come from two very different mathematical frameworks. The questions this work aims to address are the following: \textit{Can we unify the most popular SBL algorithms under a single mathematical framework? Using insights from this framework can we design data driven techniques to learn a superior SBL algorithm?} The following contributions of the paper aim to answer the above two questions. 
\begin{enumerate}
    \item We analyze the popular SBL algorithms and introduce a novel re-parameterization that allows us to express/rewrite the SBL update rules in a form suitable for subsequent analysis and enhancement. This parameterization at the outset provides a novel re-interpretation of the EM-SBL algorithm. Using the insights from the minimum power distortionless response (MPDR) view of the SBL algorithm~\cite{mpdr_sbl,pote2023maximum}, it also enables providing a beamforming interpretation to the different mappings used at each iteration by the different SBL algorithms. 
    \item Using the new parameterization, we re-derive the MU-SBL update rule using the majorization-minimization (MM) framework, where the multiplicative update rule is a consequence of majorizing the SBL objective function in a specific way. The new majorizer provides convergence guarantees for the multiplicative update rule and reveals a new class of SBL algorithms. A consequence of this majorizer is that it has unified the popular SBL algorithms under the MM framework. 
    In this work, we provide an even stronger relationship between the MU-SBL and EM-SBL update rule by showing that not only do both of them fall under the same MM framework, but both algorithms correspond to minimizing the same majorizer in different ways. The study of SBL algorithms has received much attention~\cite{wipf2011latent,wipf2007new,wipf2010iterative,ye2015improving,hashemi2021unification,peter,nannuru2019sparse} and we provide a comparison with past work in section~\ref{sec:reparam}.
    \item Using the insights provided by the new formulations, we develop a learning method to first learn the best SBL algorithm by restricting the mapping to the class of majorizers. We first showcase different ways to expand the class of majorizers and the corresponding update rules. We then introduce a novel way to learn the best update rule from the data.  
    \item We then go beyond this restrictive class to a richer class by parameterizing it with a novel neural network architecture that builds on our reparameterization of SBL. This allows us to learn a more general mapping. We show that such a model when trained on a particular type of matrices allows us to \textit{generalize to unseen matrices of different sizes as well as different sparsity levels, signal-to-noise ratios, and snapshots} achieving our desired goal of learning a superior SBL algorithm. 
\end{enumerate}

\subsection{Notation and Outline}
We denote vectors in lower-case bold letters as $\mathbf{x}$ and matrices as upper case bold letters $\boldsymbol{X}$. The absolute value operator and the determinant operator are overloaded with the same symbol $|\cdot|$, however, when used in context, it is clear which operation is being used. The absolute operator when used on a vector applies the absolute operation element-wise. In this paper, we adopt the notation of division and multiplication between two vectors as their element-wise operations. Squaring of a vector is the element wise square operation. Indexing the $i^{\text{th}}$ element of a vector $\mathbf{x}$ is denoted as $\mathbf{x}[i]$. The set $\{1,\ldots,N\}$ is denoted by $[N]$. The rest of the paper is organized as follows. Section~\ref{sec:intro} introduces the canonical form of SBL for SSR, and in section~\ref{sec:reparam} we introduce a novel reparameterization of popular SBL algorithms, as well as introduce and analyze a new class of SBL algorithms. In section~\ref{sec:mm_learn} we expand this class using principles of MM and show how one could learn the best majorizer via data. We build on these insights to develop a neural network architecture in section~\ref{sec:deep_learning} and the results are showcased in~\ref{sec:results}. 

\section{Sparse Bayesian Learning}
\label{sec:intro}
Sparse Bayesian Learning (SBL) is an empirical Bayesian method to find the sparsest solution to~\eqref{eq:inverse_problem}. It assigns a parameterized prior to $\boldsymbol{X}$ and learns the parameters through the measurements $\boldsymbol{Y}$. Typically, SBL falls under the framework of Type-II maximization problems where the maximum likelihood problem is not solved for $\boldsymbol{X}$ but rather the parameters of its prior. In this work, the prior on $\boldsymbol{X}$ is given by $\Prob{\boldsymbol{X};\boldsymbol{\gamma}}=\prod^L_{l=1}\Prob{\vecsym{x}_l;\boldsymbol{\gamma}}$, where $\boldsymbol{\gamma}\in\mathbb{R}^{M}_{+}$ with $\boldsymbol{\gamma} = [\boldsymbol{\gamma}[1], \hdots,\boldsymbol{\gamma}[M]]^T.$ The prior on each measurement is given as
\begin{equation*}
    \Prob{\vecsym{x}_l;\boldsymbol{\gamma}} = \prod^{M}_{i=1}\Prob{\vecsym{x}_l[i];\boldsymbol{\gamma}[i]} = \prod^{M}_{i=1} \mathcal{N}\left(\vecsym{x}_l[i];0,\boldsymbol{\gamma}[i]\right),
\end{equation*}
where $\mathcal{N}\left(0,\boldsymbol{\gamma}[i]\right)$ is the standard Gaussian density with mean 0 and variance $\boldsymbol{\gamma}[i].$ Note that the sparsity profile of $\vecsym{x}_l$ and $\boldsymbol{\gamma}$ are the same, that is, if $\boldsymbol{\gamma}[i]=0$ then $\vecsym{x}_l[i]=0\;\forall\; l \in [L]$. The Type-II maximum likelihood (ML) problem~\cite{wipf} in $\boldsymbol{\gamma}$-space is given as 
\begin{equation}
\label{eq:type_2_ml}
    \boldsymbol{\gamma}_* = \argmax_{\gamma \in \mathbb{R}^M_+} \Prob{\mathbf{Y}| \boldsymbol{\gamma}}.
\end{equation}
Combining the measurement model and the Gaussian prior on $\boldsymbol{X}$ results in the following ML optimization problem.
\begin{equation}
\begin{split}
    \boldsymbol{\gamma}_{*} &= \argmin_{\boldsymbol{\gamma} \in \mathbb{R}^{M}_{+}} f(\boldsymbol{\gamma}) \label{eq:nll}\\&=  \argmin_{\boldsymbol{\gamma} \in \mathbb{R}^{M}_{+}} \log|\covyy(\boldsymbol{\gamma})| + \frac{1}{L}\sum^{L}_{l=1}\vecsym{y}_l^{T}\covyy^{-1}(\boldsymbol{\gamma})\vecsym{y}_l,
    \end{split}
\end{equation}
where $\covyy(\boldsymbol{\gamma})=\mm \boldsymbol{\Gamma} \mm^{T} + \sigma^2 \mathbf{I}_N$,{\footnote{In this work, we consider $\sigma^2>0$. Hence, the covariance matrix $\covyy(\boldsymbol{\gamma})$ is positive definite.}} and $\boldsymbol{\Gamma}=\text{diag}(\boldsymbol{\gamma})$. The objective function in~\eqref{eq:nll} is the negative log marginal likelihood (up to constants), obtained by integrating out $\mathbf{X}$ from the joint distribution $p(\mathbf{Y}, \mathbf{X}| \boldsymbol{\gamma})$. If a separable prior was chosen for $\boldsymbol{\gamma}$ such that $\Prob{\boldsymbol{\gamma}}=\prod^{M}_{i=1}\Prob{\boldsymbol{\gamma}[i]}$, then the prior on $x_i$ belong to the Gaussian scale mixture family~\cite{andrews1974scale,palmer2005variational}. The MAP estimate of $\boldsymbol{\gamma}$ is very similar to~\eqref{eq:nll}, albeit with an extra term of $-2\sum^{M}_{i=1}\log(\Prob{\boldsymbol{\gamma}[i]})$, that is,
\begin{equation}
\label{MAP}
\begin{split}
    \boldsymbol{\gamma}_{MAP} &=  \arg\min_{\boldsymbol{\gamma} \in \mathbb{R}^{M}_{+}}\left[ \log|\covyy(\boldsymbol{\gamma})| + \frac{1}{L}\sum^{L}_{l=1}\vecsym{y}_l^{T}\covyy^{-1}(\boldsymbol{\gamma})\vecsym{y}_l \right. \\&\left.-2\sum^{M}_{i=1}\log(\Prob{\boldsymbol{\gamma}[i]})\right].
\end{split}\end{equation}
Though we focus on the maximum likelihood problem, some of the observations also benefit the MAP problem. Additionally, if $L>N$, then one can solve an equivalent simpler problem where the sum in~\eqref{eq:nll} has $N$ terms instead of $L$ terms~\cite{wipf}. 

The posterior density $\Prob{\boldsymbol{X}|\boldsymbol{Y},\boldsymbol{\gamma}}$ is completely specified once we find the optimal solution $\boldsymbol{\gamma}_{*}$. A point estimate for $\boldsymbol{X}$ can be that of the mean of the density $\Prob{\boldsymbol{X}|\boldsymbol{\boldsymbol{Y},\boldsymbol{\hat{\gamma}}}}$, which is a Gaussian density whose first and second order moments are given as
\begin{equation}
\begin{split}
    \boldsymbol{\hat{X}(\boldsymbol{\hat{\gamma}})} &= \boldsymbol{\hat{\Gamma}}\boldsymbol{\Phi}^{T}\boldsymbol{\Sigma}^{-1}_{yy}(\boldsymbol{\hat{\gamma}})\boldsymbol{Y},\label{eq:mean}\\
\boldsymbol{\Sigma}_e(\boldsymbol{\hat{\gamma}}) &= \boldsymbol{\hat{\Gamma}}-\boldsymbol{\hat{\Gamma}}\boldsymbol{\Phi}^{T}\boldsymbol{\Sigma}^{-1}_{yy}(\boldsymbol{\hat{\gamma}})\boldsymbol{\Phi} \boldsymbol{\hat{\Gamma}},
\end{split}
\end{equation}
where $\boldsymbol{\hat{\gamma}}$ is the estimate of the ground truth $\boldsymbol{\gamma}$, and $\boldsymbol{\hat{\Gamma}}=\text{diag}(\boldsymbol{\hat{\gamma}})$.
\subsection{Classical SBL Algorithms}
 Choices for solving the non-convex optimization problem outlined in~\eqref{eq:type_2_ml} include the EM-SBL~\cite{wipf} approach and the MU-SBL~\cite{tipping}, which are summarized next.
\subsubsection{Expectation-Maximization Algorithm (EM-SBL)}
In every iteration, the EM algorithm constructs a surrogate function which is an upper bound to~\eqref{eq:type_2_ml}, and then minimizes the surrogate to derive the update rule for the estimate of  $\boldsymbol{\gamma}$. The EM majorizer, denoted by $g_{\text{EM}}(\boldsymbol{\gamma}| \hat{\boldsymbol{\gamma}}^j)$ at $\hat{\boldsymbol{\gamma}}^j$ in the $j^{\text{th}}$ iteration was originally derived in \cite{tipping}, and can be shown to be
\begin{multline}
\label{eq:ema_maj}
    g_{\text{EM}}(\boldsymbol{\gamma};\boldsymbol{\hat{\gamma}}_{j}) = \frac{1}{2}\sum^{M}_{i=1} \Biggl( \log{\boldsymbol{\gamma}[i]} + \\ 
    \frac{\frac{1}{L}\sum^{L}_{l=1}\mathbf{\hat{x}}^2_l(\boldsymbol{\hat{\gamma}}_j)[i] + \boldsymbol{\Sigma}_e(\boldsymbol{\hat{\gamma}}_j)[i,i]}{\boldsymbol{\gamma}[i]} \Biggr) + \text{const.}
\end{multline}
The corresponding update rule obtained by minimizing the above majorizer is given as follows
\begin{equation}
    \label{eq:em_update}
    \boldsymbol{\hat{\gamma}}_{j+1}= \frac{1}{L}\sum^{L}_{l=1}\mathbf{\hat{x}}^2_l(\boldsymbol{\hat{\gamma}}_j) + \text{diag}(\boldsymbol{\Sigma}_e(\boldsymbol{\hat{\gamma}}_j)).
\end{equation}

The EM algorithm falls under a class of general algorithms referred as the majorization-minimization (MM) methods~\cite{mm_paper} and inherits its advantages mainly that of convergence to local minima of the objective function~\eqref{eq:type_2_ml}.
\subsubsection{Tipping's Multiplicative Update Rule (MU-SBL)}
The multiplicative update rule, derived in~\cite{tipping} using the approach of Mackay in \cite{mackay1992bayesian}, is obtained by setting the gradient of the objective function~\eqref{eq:type_2_ml} to zero, and then using a fixed-point approach to update the values of $\boldsymbol{\gamma}$. The multiplicative update rule  is given by~\cite{tipping}
\begin{equation}
\label{eq:tipping_update}
   \boldsymbol{\hat{\gamma}}_{j+1} = \frac{\frac{1}{L}\sum^{L}_{l=1}\mathbf{\hat{x}}^2_l(\boldsymbol{\hat{\gamma}}_j)}{\boldsymbol{\hat{\gamma}}_{j}-\text{diag}(\boldsymbol{\Sigma}_e(\boldsymbol{\hat{\gamma}}_j))}\boldsymbol{\hat{\gamma}}_j. 
\end{equation}
An interesting observation common to both update rules is that if $\boldsymbol{\hat{\gamma}}_{j}[i]=0$, then for all future iterations, the $i^{\text{th}}$ component of $\boldsymbol{\hat{\gamma}}_k,\forall \; k>j$ will be zero, implying from~\eqref{eq:mean} that $\vecsym{\hat{x}}_l(\boldsymbol{\hat{\gamma}}_k)[i]=0\;\forall\; l \in [L]$. 
\section{MM-Based SBL algorithms: Re-interpreting Classical SBL Algorithms}
\label{sec:reparam}
The EM update rule and the multiplicative update rule arise from two very different methodologies to solve~\eqref{eq:nll}. In this section, we derive a new majorizer for the objective function in~\eqref{eq:nll}, and minimizing this majorizer leads to a new class of SBL algorithms. We show that the multiplicative update rule is part of this new class. Before we move to the derivation, we introduce two key vector quantities which are defined as
\begin{align}
    \mathbf{T}_1(\boldsymbol{\gamma}) &= \frac{1}{L}\sum^L_{l=1}|\mm^{T}\covyy^{-1}(\boldsymbol{\gamma})\vecsym{y}_l|^2,\\
    \mathbf{T}_2(\boldsymbol{\gamma}) &= \text{diag}(|\mm^{T}\covyy^{-1}(\boldsymbol{\gamma})\mm|),
\end{align}
where $\mathbf{T}_1(\boldsymbol{\gamma}),\mathbf{T}_2(\boldsymbol{\gamma}) \in \mathbb{R}^{M}_{+}$. We note that related quantities appear in the fast marginal likelihood maximization algorithm in~\cite{tipping2003fast}, where $\mathbf{T}_1(\boldsymbol{\gamma})[i]$ corresponds to $Q_i^2$ and $\mathbf{T}_2(\boldsymbol{\gamma})[i]$ corresponds to $S_i$ in their notation. While these quantities were introduced therein as intermediate variables for algorithmic efficiency, we bring them to the forefront and provide a systematic interpretation through the MPDR framework~\cite{mpdr_sbl}, demonstrating their central role in unifying and extending SBL algorithms. Moreover these quantities appear in the sequential basis selection problem which can be interpreted using the Minimum Variance Distortionless Response (MVDR) framework~\cite{pote2025theory}. Given model parameters $\boldsymbol{\gamma},$ $\mathbf{T}_1(\boldsymbol{\gamma}$) is a data dependent term and $\mathbf{T}_2(\boldsymbol{\gamma}$) is a purely model dependent term. These quantities naturally arise when one takes the gradient of the  
objective function~\eqref{eq:type_2_ml} \cite{peter,ye2015improving}. We identify and bring these parameters to the forefront because of their importance in our work on unification, interpretation and enhancement of SBL algorithms.

\subsection{Re-interpreting Classical SBL Algorithms}
\subsubsection{EM-SBL~\cite{wipf}} The EM update rule at the $j^{\text{th}}$ iteration, after substituting the new quantities in~\eqref{eq:em_update}, is given by 
\begin{equation}
\label{eq:em_reint_update}
    \boldsymbol{\gamma}_{j+1} = \left(\mathbf{T}_{1}(\boldsymbol{\gamma}_j) - \mathbf{T}_{2}(\boldsymbol{\gamma}_j)\right)\boldsymbol{\gamma}^2_{j} + \boldsymbol{\gamma}_{j}.
\end{equation}
Note that the gradient of the objective function in~\eqref{eq:nll} at a point $\boldsymbol{\gamma}$ interior to the positive orthant is given by the new parameters as $\mathbf{T}_2(\boldsymbol{\gamma})-\mathbf{T}_1(\boldsymbol{\gamma})$. Given this, we can easily interpret the EM update rule as a gradient descent algorithm with adaptive step size. Interestingly, the term $(\mathbf{T}_1(\boldsymbol{\gamma}) - \mathbf{T}_2(\boldsymbol{\gamma}))$ is closely related to the relevance factor introduced in \cite{tipping2003fast}, which determines whether a basis function should be added or removed in the sequential update scheme. 
\subsubsection{MU-SBL~\cite{tipping}}
The multiplicative update rule in~\eqref{eq:tipping_update}, after some simple algebra, can be rewritten as 
\begin{equation}
\label{eq:tipping_t1_update}\boldsymbol{\gamma}_{j+1} = \frac{\mathbf{T}_1(\boldsymbol{\gamma}_j)}{\mathbf{T}_2(\boldsymbol{\gamma}_j)} \boldsymbol{\gamma}_j.
\end{equation}
To make the update rule more formally acceptable, we first provide convergence guarantees to the update rule and then introduce a novel beamforming interpretation to it.
\subsection{$p$-SBL: A new class of SBL algorithms}
In this section we derive a new family of SBL algorithms based on the MM principle, and show that the multiplicative update rule is part of it. 

\begin{theorem}
\label{th:pSBL}
The following iterative updates of $\gamma \in \mathbb{R}^{M}_{+}$ converges to a stationary point\footnote{With some perturbations introduced into the algorithm, saddle points can be avoided and likely have convergence to a local minima.} of the objective function  in~\eqref{eq:nll}
\begin{equation}
\label{eq:p_t_update}
    \boldsymbol{\gamma}_{j+1} = \left(\frac{\mathbf{T}_1(\boldsymbol{\gamma}_j)}{\mathbf{T}_2(\boldsymbol{\gamma}_j)}\right)^p\boldsymbol{\gamma}_{j},\;\forall \; 0<p\leq 1.
\end{equation}
The update rule is derived using the following majorizer of $f$
\begin{equation}
\label{eq:p_sbl_maj}
    g_{\text{$p$-SBL}}(\boldsymbol{\gamma};\boldsymbol{\hat{\gamma}})=\sum^M_{i=1}\gamma[i] \mathbf{T}_2(\boldsymbol{\hat{\gamma}})[i] + \frac{\hat{\gamma}^2[i]\mathbf{T}_1(\boldsymbol{\hat{\gamma}})[i]}{\gamma[i]}+\text{const}.
\end{equation}
Moreover, the EM update rule at the $j^{\text{th}}$ iteration given by 
    \begin{equation*}
\boldsymbol{\hat{\gamma}}^{\text{EM}}_{j+1} = \boldsymbol{\hat{\gamma}}_{\text{j}} + (\mathbf{T}_1(\boldsymbol{\hat{\gamma}}_j) - \mathbf{T}_2(\boldsymbol{\hat{\gamma}}_j))\boldsymbol{\hat{\gamma}}^2_j,
    \end{equation*}
also decreases the value of the $p$-SBL majorizer in each iteration, that is, $g_{\text{p-SBL}}(\boldsymbol{\hat{\gamma}}^{\text{EM}}_{j+1};\boldsymbol{\hat{\gamma}}_{j})\leq g_{\text{p-SBL}}(\boldsymbol{\hat{\gamma}}_{j};\boldsymbol{\hat{\gamma}}_{j}) \;\forall\;j \in \mathbb{N}$.
\begin{proof}
The proof is relegated to Appendix~\ref{app:p_sbl}.
\end{proof}
\end{theorem}

\textit{Remarks :} We make the following remarks from Theorem~\ref{th:pSBL}.
\begin{enumerate}
    \item By choosing $p=1$ the $p$-SBL update is exactly the same as MU-SBL, providing convergence proofs for the multiplicative update rule. It also provides an explanation as to why the multiplicative update rule has had performance comparable to that of EM-SBL in many scenarios, even though the fixed-point derivation of the update had no convergence guarantees provided.  
    \item The search and explanation for SBL algorithms has a rich history, and  a few related to this class of algorithms can be found in~\cite{wipf2011latent,wipf2007new,wipf2010iterative,ye2015improving,hashemi2021unification,peter,nannuru2019sparse}. The special case \(p=\tfrac12\) can be gleaned from the work in \cite{wipf2011latent,wipf2007new,mackay1992bayesian} and was explicitly stated in \cite{owen2012performance}. This was later derived in~\cite{hashemi2021unification} using a MM formulation.~\cite{peter,nannuru2019sparse} later proposed a natural generalization of their update rule to arbitrary $p$, however, without a theoretical convergence analysis.   The case of $p=1$ was discussed in Appendix E in~\cite{hashemi2021unification}, but we have difficulty with the proof. Our work for $p=1$ can be viewed as additional evidence. In this work, we cast the general-$p$ update firmly within the MM framework and provide a complete convergence guarantee for $0 < p \leq 1.$ To the best of our knowledge, these results are new.
    \item The majorizer $g_{\text{p-SBL}}$ has unified the update equations for two of the most widely used SBL algorithms. Theorem~\ref{th:pSBL} establishes that the EM update rule, while canonically derived from the $g_{\text{EM}}$ majorizer~\cite{tipping}, also constitutes a valid descent step for the $g_{\text{p-SBL}}$ majorizer. This demonstrates an analytical compatibility: both EM-SBL and MU-SBL decrease the same surrogate function $g_{\text{p-SBL}}$. The above derivation also highlights the flexibility  under the MM framework, an update rule is valid provided it guarantees descent on the majorizer---it need not be the minimizer. This insight is practically significant, as the minimizer of a majorizer is not always the most desirable update rule, as evidenced by the convergence behavior discussed in Section~\ref{sec:results}. 
\end{enumerate}
\textbf{\textit{Beamforming (BF) Interpretation}: }
The reparameterization allows one to provide insight into the algorithms using a beamforming perspective. Given model parameters $\boldsymbol{\gamma},$ $\mathbf{T}_1(\boldsymbol{\gamma}$) is a data dependent term and $\mathbf{T}_2(\boldsymbol{\gamma}$) is a purely model dependent term. Among the algorithms presented, in every iteration they check for consistency between the model and the data terms. For  given model parameters $\hat{\boldsymbol{\gamma}}_j,$ the MPDR formulation of SBL~\cite{mpdr_sbl}, which can be seen by utilizing the connection between MPDR and MMSE beamformers captured by the conditional mean in~\eqref{eq:mean} \cite{Vantrees}, provides insight. One can interpret  $\mathbf{T}_1(\boldsymbol{\hat{\gamma}})[i]/\mathbf{T}^2_2(\boldsymbol{\hat{\gamma}})[i]$ as the output of the MPDR BF's power along direction represented by column $i$, namely $\phi_i$, based on the data as input and the MPDR BF designed using the parameter estimate $\boldsymbol{\hat{\gamma}}.$ The term $1/\mathbf{T}_2(\boldsymbol{\hat{\gamma}})[i]$ is the output of the MPDR BF's  power along direction represented by $\phi_i$ based on the model alone. We see that every SBL algorithm in each iteration corrects for the source power estimates $\boldsymbol{\hat{\gamma}}_j$ for the disparity between the MPDR model power and the true power dictated by data. The only difference between the SBL algorithms is the way they define the disparity or mismatch, the EM algorithm defines it as the difference of the two powers, while the $p$-SBL algorithm takes the ratio between the two.

\section{Learning the Majorizer via  Data}
\label{sec:mm_learn}
Different majorizers for the same problem possess distinct analytical properties and often lead to different algorithms. However, it is unclear how to find the "best" majorizer for a given problem.
Given a set of majorizers and their corresponding update rules, we now present strategies to systematically generate new majorizers and update rules from existing ones which are  potentially better candidates. Since multiple update rules can minimize the same majorizer (e.g., the $p$-SBL class), the number of algorithms always exceeds or equals the number of majorizers. A similar strategy of combining different algorithms in the SSR setting has been studied in~\cite{bapat2023convex}.
\begin{definition}~\cite{lange2016mm} A function $g(x;y)$ is said to majorize a function $f:\mathbb{R}\rightarrow \mathbb{R}$ at the point $y\in \mathbb{R}$ if and only if $f(x)\leq g(x;y), \;\forall x \in \mathbb{R}$ and $g(y;y)=f(y)$.
\end{definition}
An immediate property of a majorizer is that if $g(x_{n+1};x_n)\leq g(x_n;x_n)$ then $f(x_{n+1})\leq f(x_n)$. 

\begin{definition} A valid update rule is a function $h(x_{n}) = x_{n+1}$ such that $g(x_{n+1};x_n)\leq g(x_n;x_n),\; \forall n\in \mathbb{N}$.
\end{definition}

To enable data-driven learning of optimal majorizers, we first enlarge the set of candidate majorizers and update rules. We propose two approaches: (i) generating new update rules from existing ones without altering the majorizer, and (ii) constructing new majorizers to derive novel update rules. While we focus on the objective function $f$ in~\eqref{eq:nll}, the framework extends to MAP estimation of $\boldsymbol{\gamma}$, assuming the prior term preserves the required majorizer properties.
\subsection{Convex Combination of Update Rules}If the majorizer $g(\boldsymbol{\gamma})$ has $Q$ different update rules, 
$$\boldsymbol{{\hat{\gamma}}}^{(q)}_{j}\rightarrow\boldsymbol{\hat{\gamma}}^{(q)}_{j+1}, q\in[Q],$$
then at the $j^{\text{th}}$ iteration each of the $Q$ different update rules have the following property
\begin{align*}    f(\boldsymbol{{\hat{\gamma}}}^{(q)}_{j+1})\leq g(\boldsymbol{{\hat{\gamma}}}^{(q)}_{j+1};\boldsymbol{\hat{\gamma}}^{(q)}_{j}) &\leq g(\boldsymbol{{\hat{\gamma}}}^{(q)}_{j};\boldsymbol{\hat{\gamma}}^{(q)}_{j})\\&=f(\boldsymbol{{\hat{\gamma}}}^{(q)}_{j}),\forall\;\; q\in[Q].
\end{align*}
\begin{theorem}\label{thm:convexcomb}
    If $g(\boldsymbol{\gamma})$ is a convex function and it majorizes  $f(\boldsymbol{\gamma})$, then a convex combination of its $Q$ update rules is also a valid update rule and is given by
    \begin{align*}
        \boldsymbol{\hat{\gamma}}_{j} = \sum^{Q}_{q=1} \alpha_{j,q} \boldsymbol{\hat{\gamma}}^{(q)}_{j},  
    \end{align*}
    where $\alpha_{j,q}\geq0,\;\forall q \in [Q]$ and $\sum^{Q}_{q=1}\alpha_{j,q}=1$.
    \begin{proof} The validity of the above update rule follows immediately from the convexity constraint of $g(\boldsymbol{\gamma})$, 
\begin{multline*}
f(\boldsymbol{{\hat{\gamma}}}_{j+1}) \leq g(\boldsymbol{{\hat{\gamma}}}_{j+1};\boldsymbol{\hat{\gamma}}_{j}) = g\left(\sum^{Q}_{q=1} \alpha_{j+1,q} \boldsymbol{\hat{\gamma}}^{(q)}_{j+1};\boldsymbol{\hat{\gamma}}_{j}\right) \leq \\
\sum^{Q}_{q=1} \alpha_{j+1,q} \, g(\boldsymbol{\hat{\gamma}}^{(q)}_{j+1};\boldsymbol{\hat{\gamma}}_{j}) 
\leq \sum^{Q}_{q=1} \alpha_{j+1,q} \, g(\boldsymbol{\hat{\gamma}}_{j};\boldsymbol{\hat{\gamma}}_{j}) \leq f(\boldsymbol{{\hat{\gamma}}}_{j}).
\end{multline*}
    \end{proof}
\end{theorem}
The $p$-SBL majorizer in~\eqref{eq:p_sbl_maj} is convex. Hence, from Theorem~\ref{thm:convexcomb}, a convex combination of update rules for different values of $p$ is also a valid update rule. For our application, from Theorem~\ref{th:pSBL} we see that the EM and $p$-SBL update rules correspond to the same majorizer, hence their convex combination is a valid update rule. However, the validity of the convex combination of update rules corresponding to different majorizers is still open. We outline sufficient conditions for a convex combination of update rules from different majorizers to be a valid update rule. 
\begin{theorem}
\label{th:conDiffMaj}
If the majorizers of $f(\boldsymbol{\gamma})$, $g_q(\boldsymbol{\gamma})\; \forall \;q\in\; [Q]$ are convex, and their pointwise minimum is convex, then the following is a valid update rule,
\begin{equation}
    \label{eq:convUpdate_Maj}
    \boldsymbol{\hat{\gamma}}_{j+1} = \sum^{Q}_{q=1} \sum^{S_q}_{s=1}\alpha_{q,s} \boldsymbol{\hat{\gamma}}^{q,s}_{j+1},
\end{equation}
where $\boldsymbol{\hat{\gamma}}^{q,s}_{j+1}\forall s\in [S_q]$ are the estimates obtained from the $S_q$ update rules corresponding to $g_q(\boldsymbol{\gamma};\boldsymbol{\hat{\gamma}}_{j})$ and $\alpha_{q,s}\geq 0\; \forall\; s\in [S_q], q\in [Q], \; \sum^Q_{q=1}\sum^{S_q}_{s=1}\alpha_{s,q}=1$. 
\begin{proof} The proof builds on the assumption that the following function is convex
\begin{equation}
\label{eq:min_maj}g(\boldsymbol{\gamma};\boldsymbol{\hat{\gamma}}) = \min_{i \in [Q]} g_i(\boldsymbol{\gamma};\boldsymbol{\hat{\gamma}}),
\end{equation}
and  $g(\boldsymbol{\gamma};\boldsymbol{\hat{\gamma}})$ is a majorizer of $f(\boldsymbol{\gamma})$. This is immediate because $f(\boldsymbol{\hat{\gamma}}) = g(\boldsymbol{\hat{\gamma}};\boldsymbol{\hat{\gamma}})$ and $f(\boldsymbol{\gamma})\leq g_i(\boldsymbol{\gamma};\boldsymbol{\hat{\gamma}})\;\forall i\in [Q]$, so the pointwise minimum will also obey the inequality~\cite{lange2016mm}. Using the above two properties of $g(\boldsymbol{\gamma};\boldsymbol{\hat{\gamma}})$, we derive the following
\begin{multline}
    \label{eq:minMaj_decF}f(\boldsymbol{\hat{\gamma}}_{j+1})\leq g\left(\boldsymbol{\hat{\gamma}}_{j+1};\boldsymbol{\hat{\gamma}}_j\right) = g\left(\sum^{Q}_{q=1} \sum^{S_q}_{s=1}\alpha_{q,s} \boldsymbol{\hat{\gamma}}^{q,s}_{j+1};\boldsymbol{\hat{\gamma}}_j\right)\leq  \\\left(\sum^{S_1}_{s=1}\alpha_{1,s}\right) g\left(\boldsymbol{\hat{\gamma}}^{1}_{j+1} ;\boldsymbol{\hat{\gamma}}_j\right)+\ldots+\left(\sum^{S_Q}_{s=1}\alpha_{Q,s}\right)\times \\g\left(\boldsymbol{\hat{\gamma}}^{Q}_{j+1};\boldsymbol{\hat{\gamma}}_j \right) ,
\end{multline}
where $\boldsymbol{\hat{\gamma}}^{q}_{j+1} = \frac{1}{\sum^{S_q}_{s=1}\alpha_{q,s}}\sum^{S_q}_{s=1}\alpha_{q,s}\boldsymbol{\hat{\gamma}}^{i,s}_{j+1}$. For the $q^{\text{th}}$ term in the summation above we apply the definition~\eqref{eq:min_maj} to get the following 
\begin{multline}
    \label{eq:minMaj_f}g\left(\boldsymbol{\hat{\gamma}}^{q}_{j+1} ;\boldsymbol{\hat{\gamma}}_j\right)\leq g_q\left(\boldsymbol{\hat{\gamma}}^{q}_{j+1} ;\boldsymbol{\hat{\gamma}}_j\right)\leq \\ \frac{1}{\sum^{S_q}_{s=1}\alpha_{q,s}}\sum^{S_q}_{s=1}\alpha_{q,s}g_q(\boldsymbol{\hat{\gamma}}^{q,s}_{j+1} ;\boldsymbol{\hat{\gamma}}_j) \leq \\ \frac{1}{\sum^{S_q}_{s=1}\alpha_{q,s}}\sum^{S_q}_{s=1}\alpha_{q,s}g_q(\boldsymbol{\hat{\gamma}}_{j} ;\boldsymbol{\hat{\gamma}}_j) =  f(\boldsymbol{\hat{\gamma}}_{j}).
\end{multline}
Substituting~\eqref{eq:minMaj_f} into~\eqref{eq:minMaj_decF} for each of the $Q$ terms we see that the convex combination is indeed a valid update rule. 
\end{proof}
\end{theorem}
Similar to Theorem~\ref{thm:convexcomb}, the weights used in the convex combination in Theorem~\ref{th:conDiffMaj} can be different for each iteration, however in the above theorem the dependency on $j$ is suppressed to avoid clutter. 
\subsection{Convex Combination of Majorizers}
From the theory of MM, we know that if $g_q(\boldsymbol{\gamma};\boldsymbol{\hat{\gamma}}_j), \forall\; q\in [Q]$ are majorizers of the function $f(\boldsymbol{\gamma})$, then any convex combination of the majorizers will also be a majorizer of $f(\boldsymbol{\gamma})$. The following observations elucidate this point. At the $j^{\text{th}}$ iteration we have
\begin{align*}
\begin{split}
    f(\boldsymbol{\gamma}) &\leq g_q(\boldsymbol{\gamma };\boldsymbol{\hat{\gamma}}_{j}), \\
    f(\boldsymbol{\hat{\gamma}}_{j}) &= g_q(\boldsymbol{\hat{\gamma}}_{j};\boldsymbol{\hat{\gamma}}_j), \; \forall \; q \in [Q].
\end{split}
\end{align*}
Let $\alpha_1,\alpha_2,\ldots,\alpha_Q$ be constants such that $\alpha_q\geq 0,\forall q \in[Q]$ and $\sum^{Q}_{q=1}\alpha_q =1$. Then we have that 
\begin{equation*}
\begin{split}
     f(\boldsymbol{\gamma}) &\leq \sum^{Q}_{q=1}\alpha_q g_q  (\boldsymbol{\gamma}; \boldsymbol{\hat{\gamma}}_j),\\
      f(\boldsymbol{\hat{\gamma}}_{j}) &= \sum^{Q}_{q=1}\alpha_q g_q(\boldsymbol{\hat{\gamma}}_{j};\boldsymbol{\hat{\gamma}}_{j}).
\end{split}
\end{equation*}
Hence, using the existing $Q$ majorizers of $f(\boldsymbol{\gamma})$ we derive new majorizers by taking a convex combination of them. If the majorizers are convex and  differentiable then the combined majorizer is also convex and differentiable, by setting the derivative of the combined majorizers to zero, we can derive new update rules. 

As an illustration, we apply this to the SBL objective function by combining the EM majorizer in~\eqref{eq:ema_maj} and $p$-SBL majorizer in~\eqref{eq:p_sbl_maj}. Let $\alpha_1 \geq 0$, $\alpha_2\geq 0$ such that $\alpha_1+\alpha_2 =1$, then the new majorizer is given as 
\begin{align}
\label{eq:new_maj}
g(\boldsymbol{\gamma};\boldsymbol{\hat{\gamma}}) &=  \alpha_1 g_{\text{EM}}(\boldsymbol{\gamma};\boldsymbol{\hat{\gamma}}) + \alpha_2 g_{p\text{-SBL}} (\boldsymbol{\gamma};\boldsymbol{\hat{\gamma}}).
\end{align}
Re-writing~\eqref{eq:ema_maj} in terms of $\mathbf{T}_1(\boldsymbol{\gamma})$ and $\mathbf{T}_2(\boldsymbol{\gamma})$, we substitute it in~\eqref{eq:new_maj} to get the new majorizer. We clarify that while Theorem~\ref{th:pSBL} establishes that both EM-SBL and MU-SBL are valid descent steps for the $g_{\text{p-SBL}}$ majorizer, the majorizer $g_{\text{EM}}(\boldsymbol{\gamma};\boldsymbol{\hat{\gamma}})$ corresponds to~\eqref{eq:ema_maj} and the minimizer of this majorizer is the EM update rule. The majorizer $g(\boldsymbol{\gamma}; \hat{\boldsymbol{\gamma}})$ resulting from the combination of the two majorizer has different curvature properties than either constituent, potentially yielding update rules with different convergence characteristics. This flexibility  motivates data-driven optimization over the space of majorizers and is expanded on in the next section. Taking the gradient of $g(\boldsymbol{\gamma};\boldsymbol{\hat{\gamma}})$ and setting it zero, we get the following update rule 
\begin{align}
\label{eq:convMaj_updateRule}
\begin{split}
\boldsymbol{\hat{\gamma}}_{j+1} &= 
\frac{
    \alpha_1 \boldsymbol{\hat{\gamma}}_{\text{EM},j+1}
    + 2\alpha_2 \boldsymbol{\hat{\gamma}}_j^2 \mathbf{T}_1(\boldsymbol{\hat{\gamma}}_j)
}{
    \frac{\alpha_1}{2} + \sqrt{
        \frac{\alpha_1^2}{4} + \epsilon
    }
}, \\
\epsilon&=4\alpha_2 \mathbf{T}_2(\boldsymbol{\hat{\gamma}}_j) \bigg( \frac{\alpha_1 \boldsymbol{\hat{\gamma}}_{\text{EM},j+1}}{2} + \alpha_2 \boldsymbol{\hat{\gamma}}_j^2 \mathbf{T}_1(\boldsymbol{\hat{\gamma}}_j) \bigg),
\end{split}
\end{align}
where $\boldsymbol{\hat{\gamma}}_{\text{EM},j+1}$ refers to EM update rule at the ${j+1}^{\text{th}}$ iteration given by~\eqref{eq:em_reint_update}. Note that by setting $\alpha_1=0$ we get the $p$-SBL update rule with $p=1/2,$ and by setting $\alpha_1=1$ we get the EM update rule.
\subsection{Learning an Optimal Convex Combination}
\label{sec:conv_maj}
Given the expanded class of SBL algorithms, our goal is to learn an improved algorithm tailored to a specific application. To this end, provided a training dataset and a loss metric, we optimize the weights of a convex combination of the update rules via gradient descent based techniques. Since the weights are constrained to lie in a polytope, such constrained problem cannot be solved for by autograd tools (e.g. PyTorch). Hence, we transform the constrained optimization into an unconstrained problem, allowing us to use the powerful tooling of autograd. This transformation is based on the observation that a feasible solution can be obtained by passing unconstrained learnable parameters through a softmax layer. Valid update rules can then be obtained by using the outputs of the softmax layer as the weights of the convex combination. Since the entire transformation is differentiable, we can optimize them using the provided dataset and loss metric. The experimental results as well as the training methodology are presented in Section~\ref{sec:results}. 


\section{SBL-Inspired Neural Network Architecture}
\label{sec:deep_learning}
Neural networks are attractive for our problem because of their ability to represent more general nonlinear mappings and because of their ability to take advantage of and learn  priors from the data. However, instead of training a general neural network, our approach is to exploit the structure of the problem to simplify and develop a more explainable neural architecture. 

To that end,
an insight from the above reparameterization is that the EM and $p$-SBL algorithms combine $\mathbf{T}_1(\boldsymbol{\hat{\gamma}})$, $\mathbf{T}_2(\boldsymbol{\hat{\gamma}})$ and $\boldsymbol{\hat{\gamma}}$ in different ways to obtain the next estimate for $\boldsymbol{\hat{\gamma}}$. Given this observation, the question of \textit{which is the best algorithm?} becomes \textit{what is the best mapping from $(\mathbf{T}_1({\boldsymbol{\hat{\gamma}}}_j), \mathbf{T}_2({\boldsymbol{\hat{\gamma}}}_j),{\boldsymbol{\hat{\gamma}}}_j)\rightarrow \boldsymbol{\hat{\gamma}}_{j+1}$?} Using the fact that neural networks are powerful function approximators \cite{cybenko1989approximation,chen2022improved}, we explore deep learning methods to address the question.
 We develop on the ideas of deep unrolling~\cite{deep_unroll}, where the number of iterations $J$ is fixed and in each iteration learnable parameters are introduced in a structured way to learn the mapping. 

Note that the mapping used by EM and $p$-SBL from $\left(\mathbf{T}_1({\boldsymbol{\hat{\gamma}}}_j)[i], \mathbf{T}_2({\boldsymbol{\hat{\gamma}}}_j)[i],{\boldsymbol{\hat{\gamma}}}_j[i]\right)$ to $\boldsymbol{\hat{\gamma}}_{j+1}[i]$, where $i$ represents the index of the elements of the vector, is fixed and identical $\forall\;i\in [M]$ and represents a specific handcrafted choice. From this observation we design our network such that the function in the $j^{\text{th}}$ iteration denoted by $h_{\theta_j}: (\mathbf{T}_1({\boldsymbol{\hat{\gamma}}}_j)[i],\mathbf{T}_2({\boldsymbol{\hat{\gamma}}}_j)[i],\hat{\boldsymbol{\gamma}}_{j}[i])\rightarrow \boldsymbol{\hat{\gamma}}_{j+1}[i]$, where $\theta_j$ denotes the learnable parameters in the $j^{\text{th}}$ iteration, is the {\bf{same}} $\forall\;i\in [M]$ and selected from a bigger class through learning. As shown later, this sharing network of parameters allows our model architecture to become agnostic to the size of the sensing matrix. The overall architecture of the proposed method can be represented as follows
\begin{equation*}
    \boldsymbol{\hat{\gamma}}_{j+1}[i] = h(\boldsymbol{\hat{\gamma}}_{j}[i],\mathbf{T}_1(\boldsymbol{\hat{\gamma}}_j)[i],\mathbf{T}_2(\boldsymbol{\hat{\gamma}}_j)[i];\theta_j),\forall i\in[M],
\end{equation*}
where $\theta=\{\theta_1,\theta_2,\ldots,\theta_J\}$ is the set of learnable parameters for our model. Note that each entry of $\boldsymbol{\gamma}$
is the output of a network with only three inputs.
Given that we have independent parameters in each iteration, we have provided the network with the freedom to learn different mappings in each iteration. This added degree of freedom enables the network to partially compensate for the limitations imposed by a finite number of layers (or iterations). 

Another key advantage is that training the network enables it to embed dataset-specific priors directly into the learned mapping.  In theory, these priors can be incorporated into the SBL cost function and a MAP estimate of $\boldsymbol{\gamma}$ can be solved for~\cite{palmer2005variational} as discussed in section~\ref{sec:intro}. However,  the prior for the hyperparameter is usually unknown and even if it known finding a tractable choice for optimization poses additional challenges. Hence, learning from data greatly alleviates this problem, with the network acting as a meta-learner of SBL algorithms, which otherwise would have been carefully picked by the designer via the hyperprior choice $p(\boldsymbol{\gamma})$.
\subsection{Network Architecture}
\begin{figure}
\centering
\begin{tikzpicture}

        \fill[blue!60!yellow, rounded corners] (-10,3.5) -- (-9.5,4)-- (-9.5,1)-- (-10,1.5) node[align = center,scale=0.6] (A) at (-9.7,0.7) {\textcolor{black}{Projection}\\ \textcolor{black}{Layer}};

        \draw[>=stealth,->, very thick] (-9.5,2.5) -- (-10+1.1,2.5);
        \draw[very thick] (-9.3,2.5) -- (-9.3,4.5);
        \draw[very thick] (-9.3,4.5) -- (-9.5+1.1*2+0.2,4.5);
        \draw[very thick] (-9.5+1.1*2+0.2,4.5) -- (-9.5+1.1*2+0.2,2.5);
        \fill[red] (-9.5+1.1*2+0.2,2.5) circle (0.08);

        \draw[>=stealth,->, very thick] (-10.35,3.3) -- (-10,3.3) node[scale=0.8] at (-11,3.3){ \textcolor{black}{$\mathbf{T}_1(\hat{\boldsymbol{\gamma}}_{j})[i]$}};

        \draw[>=stealth,->, very thick] (-10.35,2.5) -- (-10,2.5) node[scale=0.8] at (-11,2.5){\textcolor{black}{$\mathbf{T}_2(\hat{\boldsymbol{\gamma}}_{j})[i]$}};
    
        \draw[>=stealth,->, very thick] (-10.5,1.7) -- (-10,1.7) node[scale=0.8] at (-11,1.7){\textcolor{black}{$\hat{\boldsymbol{\gamma}}_{j}[i]$}};

        \fill[blue!60!yellow, rounded corners=2] (-10+1.1,4) rectangle (-9.5+1.1,1) node[align = center, scale=0.6] (A) at (-9.7+1.1,0.7) {\textcolor{black}{Layer} \\ \textcolor{black}{1}};
        \fill[black!60!yellow] (-9.5+1.1,4) rectangle (-9.5+1.2,1);
        \draw[>=stealth,->, very thick] (-9.5+1.2,2.5) -- (-10+1.1*2,2.5);

        \fill[blue!60!yellow, rounded corners=2] (-10+1.1*2,4) rectangle (-9.5+1.1*2,1) node[align = center, scale=0.6] (A) at (-9.7+1.1*2,0.7) {\textcolor{black}{Layer} \\ \textcolor{black}{2}};
        \fill[black!60!yellow] (-9.5+1.1*2,4) rectangle (-9.5+1.1*2+0.1,1);
        \draw[>=stealth,->, very thick] (-9.5+1.1*2+0.1,2.5) -- (-10+1.1*3,2.5);

        \draw[very thick] (-9.5+1.1*2+0.38,2.5) -- (-9.5+1.1*2+0.38,4.5);
        \draw[very thick] (-9.5+1.1*2+0.38,4.5) -- (-9.5+1.1*3+0.25,4.5);
        \draw[very thick] (-9.5+1.1*3+0.25,4.5) -- (-9.5+1.1*3+0.25,2.5);
        \fill[red] (-9.5+1.1*3+0.25,2.5) circle (0.08);

        \fill[blue!60!yellow, rounded corners=2] (-10+1.1*3,4) rectangle (-9.5+1.1*3,1) node[align = center, scale=0.6] (A) at (-9.7+1.1*3,0.7) {\textcolor{black}{Layer} \\ \textcolor{black}{3}};
        \fill[black!60!yellow] (-9.5+1.1*3,4) rectangle (-9.5+1.1*3+0.1,1);
        \draw[>=stealth,->, very thick] (-9.5+1.1*3+0.1,2.5) -- (-10+1.1*4,2.5);
        
        \fill[blue!60!yellow, rounded corners=2] (-10+1.1*4,4) rectangle (-9.5+1.1*4,1) node[align = center, scale=0.6] (A) at (-9.7+1.1*4,0.7) {\textcolor{black}{Layer} \\ \textcolor{black}{4}};
        \fill[black!60!yellow] (-9.5+1.1*4,4) rectangle (-9.5+1.1*4+0.1,1);
        \draw[>=stealth,->, very thick] (-9.5+1.1*4+0.1,2.5) -- (-10+1.1*5,2.5);

        \fill[blue!60!yellow, rounded corners=5] (-10+1.1*5,4) -- (-9.5+1.1*5,3)-- (-9.5+1.1*5,2) --(-10+1.1*5,1) node[align = center, scale=0.6] (A) at (-9.7+1.1*5,0.7) {\textcolor{black}{Output} \\ \textcolor{black}{Layer}};
        \fill[black!60!red] (-9.5+1.1*5,3) rectangle (-9.5+1.1*5+0.1,2);
        \draw[>=stealth,->, very thick] (-9.5+1.1*5+0.1,2.5) -- (-9.5+1.1*5+0.4,2.5) node[scale=0.8] at (-9.5+1.1*5+0.8,2.55){\textcolor{black}{$\boldsymbol{\hat{\gamma}}_{j+1}[i]$}};

        \draw[very thick] (-10.25,1.7) -- (-10.25,5.2); 
        \draw[very thick] (-10.25,5.2) -- (-9.5+1.1*5+0.1+0.1,5.2); 
        \draw[very thick] (-9.5+1.1*5+0.1+0.1,5.2) -- (-9.5+1.1*5+0.1+0.1,2.5); 
        \fill[red] (-9.5+1.1*5+0.1+0.1,2.5) circle (0.08); 
        \node at (-10.25+3,5.4) {\small$\boldsymbol{g(\cdot)}$}; 
        
    \end{tikzpicture}
    
    \caption{The architecture of the $j^{\text{th}}$ iteration. It consists of 6 layers with skip connections. The red dots represent the summation operation. }
    \label{fig:iter_arch}
\end{figure}
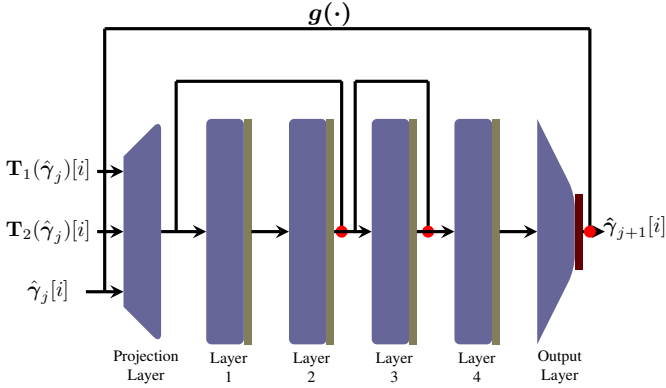

The network consists of $J$ iterations. Fig.~\ref{fig:nn_arch} shows the architecture of our neural network model. Fig.~\ref{fig:iter_arch} shows the architecture of one iteration in our neural network model. Each iteration is a multi-layer perceptron (MLP) with 4 hidden layers, each with dimension $d$, and the activation function is given by the rectified linear unit (ReLU) illustrated by the gray blocks, the input to each iteration is a vector of size 3 and the output is a scalar value. To ensure that the output is non-negative we use the softplus~\cite{softplus} activation function at the output neuron, illustrated by the red block in the figure. This function is similar to ReLU but it is differentiable at the origin. We choose this activation function because the values of $\boldsymbol{\gamma}$ close to zero are important, since these values determine whether that component of the vector is zero or of small magnitude. More importantly, during training when a $\boldsymbol{\hat{\gamma}}_j[i]$ is wrongly driven towards zero, we would like it to quickly recover to a non-zero value. In this regime it is critical for the gradients to flow back smoothly to the previous layers when $\boldsymbol{\hat{\gamma}}_j[i]\approx 0$. The softplus activation function allows for this.

Note that our goal is to learn an update rule. By sharing the weights of each iteration, that is by using the same mapping to generate all the components of $\boldsymbol{\hat{\gamma}}_j$ we achieve an advantage, enjoyed by classical algorithms, which is that the network is independent of the size of the measurement matrix $\mm$. We achieve this independence  because $\mathbf{T}_1(\boldsymbol{\gamma})$ and $\mathbf{T}_2(\boldsymbol{\gamma})$ captures all the necessary information about $\mm$ for SBL to recover the support. This makes our network standout because it provides us a unique opportunity to train our model on a particular measurement matrix, say $\mm_1$, but test it on a different measurement matrix $\mm_2$. The question of generalization to unseen measurements $\boldsymbol{Y}$ for a given $\mm$, as well as to unseen measurement matrices are discussed in more detail in section~\ref{sec:results}.

\textit{Remark: }The network architecture requires the precise knowledge of the noise variance. However, we outline possible ways one could address this issue if it were not known. It can be estimated separately using various signal processing based techniques~\cite{wipf}. Given that EM-SBL performs better when the variance used by the algorithm is a scaled version of the true variance~\cite{srikrishnan2018addressing}, the estimation of the variance  could be integrated into the network architecture. A simple modification could be of regressing the variance estimate using $\mathbf{T}_1(\boldsymbol{\gamma})$ since it has been shown that the variance estimation depends on the posterior mean~\cite{palmer2005variational}. 

\begin{figure}

\centering
\begin {tikzpicture}
    \fill[blue!40!white, rounded corners] (-8,4) rectangle (-7.3,1) node[align = center] (A) at (-7.5,0.7) {\textcolor{black}{Iteration 1}};
    \draw[>=stealth,->, very thick] (-8.5,3.3) -- (-8,3.3) node[scale=0.9] at (-9.3,3.3){\textcolor{black}{$\mathbf{T}_1(\boldsymbol{\hat{\gamma}}_0)[i]$}};

    \draw[>=stealth,->, very thick] (-8.5,2.5) -- (-8,2.5) node[scale=0.9] at (-9.3,2.5){ \textcolor{black}{$\mathbf{T}_2(\boldsymbol{\hat{\gamma}}_0)[i]$}};

    \draw[>=stealth,->, very thick] (-8.5,1.7) -- (-8,1.7) node[scale=0.9] at (-9,1.7){\textcolor{black}{$\boldsymbol{\hat{\gamma}}_0[i]$}};


    \draw[>=stealth,->, very thick] (-7.3,1.7) -- (-5.6,1.7) node[scale=0.9] at (-6.2,2){\textcolor{black}{$\boldsymbol{\hat{\gamma}}_1[i]$}};

    \draw[dashed,->,very thick] (-7.2,1.7) -- (-7.2,5) node[scale=0.9] at (-7.5,4.7){\textcolor{black}{$\boldsymbol{\hat{\gamma}}_1$}};

    \fill[yellow!70!black, rounded corners] (-7.1-0.9-0.5,6) rectangle (-5.5-1,5) node[scale=0.9,align = center] (B) at (-6.3-0.68-0.5,5.5) {\textcolor{black}{$\mathbf{T}_1 | \mathbf{T}_2$}\\ \textcolor{black}{block}};
    \draw[>=stealth,->, very thick] (-8.5-0.5,5.7+0.1) -- (-8-0.5,5.7+0.1) node[] at (-8.7-0.5,5.7+0.1){ \textcolor{black}{$\boldsymbol{\Phi}$}};
    \draw[>=stealth,->, very thick] (-8.5-0.5,5.3+0.18) -- (-8-0.5,5.3+0.18) node[] at (-8.7-0.5,5.3+0.18){ \textcolor{black}{$\boldsymbol{Y}$}};
    \draw[>=stealth,->, very thick] (-8.5-0.5,5.2) -- (-8-0.5,5.2) node[] at (-8.7-0.5,5.2){ \textcolor{black}{$\sigma^2$}};

    \draw[-,very thick] (-6.9,5) -- (-6.9,2.5) ;

    \fill[red!40!white, rounded corners] (-5.6,4) rectangle (-4.9,1)  node[align = center] (A) at (-5.2,0.7) {\textcolor{black}{Iteration 2}};
    
    \draw[>=stealth,->, very thick] (-6.9,3.4) -- (-5.6,3.4) node[scale=0.75] at (-6.22,3.7){ \textcolor{black}{$\mathbf{T}_1(\boldsymbol{\hat{\gamma}}_1)[i]$}};
    \draw[>=stealth,->, very thick] (-6.9,2.5) -- (-5.6,2.5) node[scale=0.75] at (-6.22,2.8){ \textcolor{black}{$\mathbf{T}_2(\boldsymbol{\hat{\gamma}}_1)[i]$}};

    \draw[>=stealth,->, very thick] (-4.9,1.7) -- (-3.8,1.7) node[scale=0.9] at (-3.9,2){\textcolor{black}{$\boldsymbol{\hat{\gamma}}_2[i]$}};

    \draw[dashed,->,very thick] (-4.8,1.7) -- (-4.8,5) node[scale=0.9] at (-5.1,4.7){\textcolor{black}{$\boldsymbol{\hat{\gamma}}_2$}};
    
    \fill[yellow!70!black, rounded corners] (-5.5,6) rectangle (-3.8,5) node[align = center] (B) at (-4.65,5.5) {\textcolor{black}{$\mathbf{T}_1 | \mathbf{T}_2$}\\\textcolor{black}{block}};
    \draw[>=stealth,->, very thick] (-5.8,5.7+0.1) -- (-5.5,5.7+0.1) node[] at (-6,5.7+0.1){ \textcolor{black}{$\boldsymbol{\Phi}$}};
    \draw[>=stealth,->, very thick] (-5.8,5.3+0.18) -- (-5.5,5.3+0.18) node[] at (-6,5.3+0.18){ \textcolor{black}{$\boldsymbol{Y}$}};
    \draw[>=stealth,->, very thick] (-5.8,5.2) -- (-5.5,5.2) node[] at (-6,5.2){ \textcolor{black}{$\sigma^2$}};

    \draw[-,very thick] (-4.45,5) -- (-4.45,2.5) ;

    \draw[>=stealth,->, very thick] (-4.45,3.4) -- (-3.8,3.4) node[scale=0.75] at (-3.75,3.7){ \textcolor{black}{$\mathbf{T}_1(\boldsymbol{\hat{\gamma}}_2)[i]$}};
    \draw[>=stealth,->, very thick] (-4.45,2.5) -- (-3.8,2.5) node[scale=0.75] at (-3.75,2.8){ \textcolor{black}{$\mathbf{T}_2(\boldsymbol{\hat{\gamma}}_2)[i]$}};

    \draw[loosely dotted,very thick] (-3.6,2.5) -- (-2.4,2.5) ;

    \fill[ green!70!black, rounded corners] (-2.5,4) rectangle (-1.8,1) node[align = center] (A) at (-2.1,0.7) {\textcolor{black}{Iteration $J$}};
    
    \draw[>=stealth,->, very thick, dashed] (-1.8,2.5) -- (-1.5,2.5) node[scale=0.9] at (-1.2,2.6){ \textcolor{black}{$\hat{\boldsymbol{\gamma}}_J$}};
    
\end{tikzpicture}
\caption{Neural network architecture for the learning new SBL algorithms.}
\label{fig:nn_arch}
\end{figure}
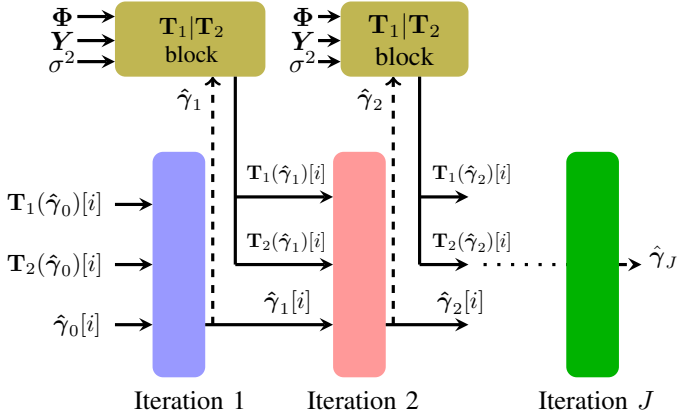

\subsection{SBL-Residual Learning}
Residual connections~\cite{he2016deep,chen2021resnests} have made training very deep models easier by allowing the gradients to flow all the way back to the input. Inside an iteration block, we introduce the identity skip connections to leverage this advantage. However, the skip connection when employed across iterations in our case does not serve a useful purpose and nor does it incorporate prior knowledge. Hence, in order to maintain the evolution of $\hat{\boldsymbol{\gamma}}_0$ to $\hat{\boldsymbol{\gamma}}_J$, and also benefit from the numerous advantages of a skip connection we incorporate one of the popular SBL update rules into the skip connection. The DNN-SBL model along with the skip connection in the $j^{\text{th}}$ iteration can be modelled as
\begin{multline*}
    \boldsymbol{\gamma}_{j+1}[i] = h(\boldsymbol{\hat{\gamma}}_{j}[i],\mathbf{T}_1(\boldsymbol{\hat{\gamma}}_j)[i],\mathbf{T}_2(\boldsymbol{\hat{\gamma}}_j)[i];\theta_j) \\+ g\left(\mathbf{T}_1(\hat{\boldsymbol{\gamma}}_j)[i],\mathbf{T}_2(\hat{\boldsymbol{\gamma}}_j)[i],\hat{\boldsymbol{\gamma}}_{j}[i]\right),\forall i\in[M],
\end{multline*}
Here, $h(\cdot~;\theta_j)$ is a function which is parameterized by the weights of a neural network and $g$ is a fixed mapping which can be any of the functions~\eqref{eq:em_reint_update},~\eqref{eq:tipping_t1_update}, or~\eqref{eq:p_t_update}. Given the choices for the update rule used in the skip connection, we consider a convex combination of the update rules and learn the weights jointly with the network. The weights for the convex combination are obtained using the same strategy as in section~\ref{sec:conv_maj}. An important benefit that arises by incorporating these prior update rules into the skip connection is that it implicitly biases the value of the original non-convex function to decreases in each iteration. While~\cite{hadou2024robust} encourages a monotonic decrease in the objective function by incorporating the difference in gradient norms across iterations into the loss function, it does not explicitly enforce descent direction through a constraint. We introduce an inductive bias towards learning such an update rule. 
\subsection{Loss Functions}
The loss function we consider while training is the exponentially weighted mean square error (MSE) loss. At each iteration we calculate the posterior mean $\boldsymbol{\hat{X}}(\boldsymbol{\hat{\gamma}}_j)$ using~\eqref{eq:mean}, and use the MSE loss as the loss criterion. The MSE loss is given by
\begin{equation*}
    \mathcal{L}_{w\text{MSE}} \triangleq \sum^{J}_{j=1}w_j ||\boldsymbol{X}(\boldsymbol{\gamma}_{*})-\boldsymbol{\hat{X}}(\boldsymbol{\hat{\gamma}}_j)||^2_F,
\end{equation*}
where $\boldsymbol{X}(\boldsymbol{\gamma}_{*})$ is the ground truth and $w_j= c^{J-j}$ where $c$ is a constant such that $c\in (0,1]$. As a proxy to the support recovery loss, we introduce the cross-entropy loss in addition to the MSE loss. We concatenate the outputs from each iteration into a vector and then pass it through a neural network (NN). In this work, we choose a MLP with ReLU non-linearity, and the final output is obtained by applying a softmax layer. The target vector is a binary vector in $\vecsym{t}_{*}\in \{0,1\}^M$ with non-zero values corresponding to the support of $\boldsymbol{\hat{X}}_*$. The final loss function is the weighted combination of the two metrics and is given as
\begin{equation}
\label{eqn:lossfunc}
    \mathcal{L} = k_1\mathcal{L}_{w\text{MSE}}(\vecsym{x_{*}}) + k_2\mathcal{L}_{\text{CE}},
\end{equation}
where the cross-entropy loss is given by
\begin{align}
\begin{split}
    \mathcal{L}_{\text{CE}} &\triangleq -\sum^{M}_{i=1}\vecsym{t}_{*}[i]\log(\hat{\vecsym{t}}[i]),\\
    \hat{\vecsym{t}} &= \text{Softmax}(\text{NN}(\boldsymbol{\hat{\gamma}}_a)),
\end{split}
\end{align}
where $\boldsymbol{\hat{\gamma}}_a =[\boldsymbol{\hat{\gamma}}^T_1,\ldots,\boldsymbol{\hat{\gamma}}^T_J]^T$. In this work, we choose $k_1=k_2=1$. Note that the NN defined in the cross-entropy loss is discarded at inference, and is used only as a means to provide gradient information about the support recovery loss to all the parameters in the network. 
\section{Results}
\label{sec:results}  
In this section we highlight the generalization capabilities of our network across snapshots, different SNRs, different measurement matrices. At this point, it is useful to note that though the exposition of the theory until now has been restricted to the field of real numbers, the SBL algorithms and, by association, the MM theory can be easily extended to the field of complex numbers~\cite{wipf}. Note that our neural network model takes as input $\mathbf{T}_1(\boldsymbol{\gamma}),\mathbf{T}_2(\boldsymbol{\gamma}),\boldsymbol{\gamma}$, all of which are in the real field irrespective of the real or complex fields of the underlying measurements or sensing matrix. Hence, no additional architectural changes are required to handle the complex field. For the training of the neural network, a  dataset is required which is described next. 

The dataset in our framework is defined as $\mathcal{T}_{\mm} = \{((\mm,\vecsym{y}_1),\vecsym{x}_1),((\mm,\vecsym{y}_2),\vecsym{x}_2),\ldots\}$, where $|\mathcal{T}_{\mm}|$ is the total number of training examples. Since our model architecture is not directly influenced by $\mm$, we also train it on a dataset consisting of different measurement matrices of different sizes which is denoted by $\mathcal{T}=\{((\mm_1,\vecsym{y}_1),\vecsym{x}_1),((\mm_2,\vecsym{y}_2),\vecsym{x}_2),\ldots\}$. For our experiments, we consider three different types of measurement matrices 
\begin{enumerate}
    \item Complex random matrix ($\boldsymbol{\Phi}_{R}$): The real and imaginary parts of each element of the matrix are sampled independently and identically from the standard Gaussian distribution. 
    \item Uniform linear array (ULA) manifold matrix: A Vandermode matrix denoted by $\boldsymbol{\Phi}_A = [\phi_1,\ldots,\phi_M]$. Each column is defined as follows
    \begin{equation*}
        \phi_i = \left[1,e^{i\pi \cos(\beta_i)},\ldots,e^{i\pi(N-1) \cos(\beta_i)}\right]^T,
    \end{equation*}
    where $\beta_1,\ldots,\beta_M$ are sampled from a maximum interval of $[0,\pi]$.
    \item Correlated matrix: It is given by $
        \boldsymbol{\Phi}_C = \sum^{N}_{i=1} \frac{1}{i^2}\mathbf{u}_i \mathbf{v}^{T}_i$,
    where $\mathbf{u}_i\in \mathbb{R}^{N}, \mathbf{v}_i\in \mathbb{R}^{M} $, and the elements of $\mathbf{u}$ and $\mathbf{v}$ are uniformly sampled from the interval $[0,1]$~~\cite{wipf_lstm}.
\end{enumerate}
 The dataset is generated using a single type of sensing matrix. Samples for both the array matrix type and the correlated matrix type are derived from a fixed sensing matrix. However, for the random matrix type, the dataset consists of different $\boldsymbol{\Phi}_R$'s, all of which are sampled from the standard Gaussian distribution. We ensure that the columns are of unit norm for all the matrices. 

The measurements $\{\boldsymbol{Y}_1,\boldsymbol{Y}_2,\ldots\}$ are generated according to the model described in~\eqref{eq:mm_model}, using one of the aforementioned types of sensing matrices and the corresponding ground truth signals (labels) $\{\boldsymbol{X}_1,\boldsymbol{X}_2,\ldots\}$ whose generation is outlined next. For each $s\in \left[\lfloor N/2 \rfloor\right]$, we generate a support set of cardinality $s$ by uniformly sampling indices from the set $[N]$. The sparsity level $s$ is constrained to be at most $N/2$, as recovery of signals with a higher sparsity level becomes fundamentally ill-posed~\cite{foucart2013mathematical}. The non-zero entries of each row of $\boldsymbol{X}_i$, corresponding to its support set are sampled i.i.d from the distribution $\mathcal{N}(0,\sigma^2_\text{s})$, where $\sigma^2_\text{s}$ denotes the power of the signal determined by the SNR. We fix the noise variance $\sigma^2=10^{-3}$ and generate the training dataset for different SNR values. The above data generation process is illustrated by fixing the number of snapshots $L$, we repeat the same process for different values of $L$. Hence, the number $|\mathcal{T}_{\boldsymbol{\Phi}}|$ enumerates all the training examples generated from different sparsity levels, SNR, and number of snapshots for a particular type of measurement matrix. 
 
 In our experiments, we benchmark against $p$-SBL and EM-SBL. The stopping criterion for EM-SBL and $p$-SBL is  $||\hat{\boldsymbol{\gamma}}_j-\boldsymbol{\hat{\gamma}}_{j+1}||_2/||\boldsymbol{\hat{\gamma}}_j||_2\leq10^{-6}$ or a maximum of 500 iterations.  To prevent premature termination, the criterion is enforced only after the first $10$ iterations. The evaluation of the SSR algorithms is in terms of two metric, 1) mean-square error and 2) the probability of support recovery. The probability of support recovery is defined as 
$    \text{PSR} = \sum^{|\mathcal{T}|}_{i=1}\mathbf{1}\{S(\boldsymbol{\gamma}_{*,i}) = S(\boldsymbol{\hat{\gamma}}_{i})\} / |\mathcal{T}|$,
where $S(\boldsymbol{\gamma}_i)$ and $S(\boldsymbol{\hat{\gamma}}_i)$ is the support set of the ground truth vector  and the predicted vector, respectively. Assuming $s=|S(\boldsymbol{\gamma}_i)|$ is known, if the predicted vector $\boldsymbol{\hat{\gamma}}$ has more than $s$ non-zero elements, the indices corresponding to the top-$s$ values of $\boldsymbol{\hat{\gamma}}_i$ becomes its support set. 
\subsection{Classical-SBL algorithms}

\begin{figure}
    \centering
    \begin{subfigure}[b]{0.4\textwidth}
    \includegraphics[height=4.5cm,width=8cm]{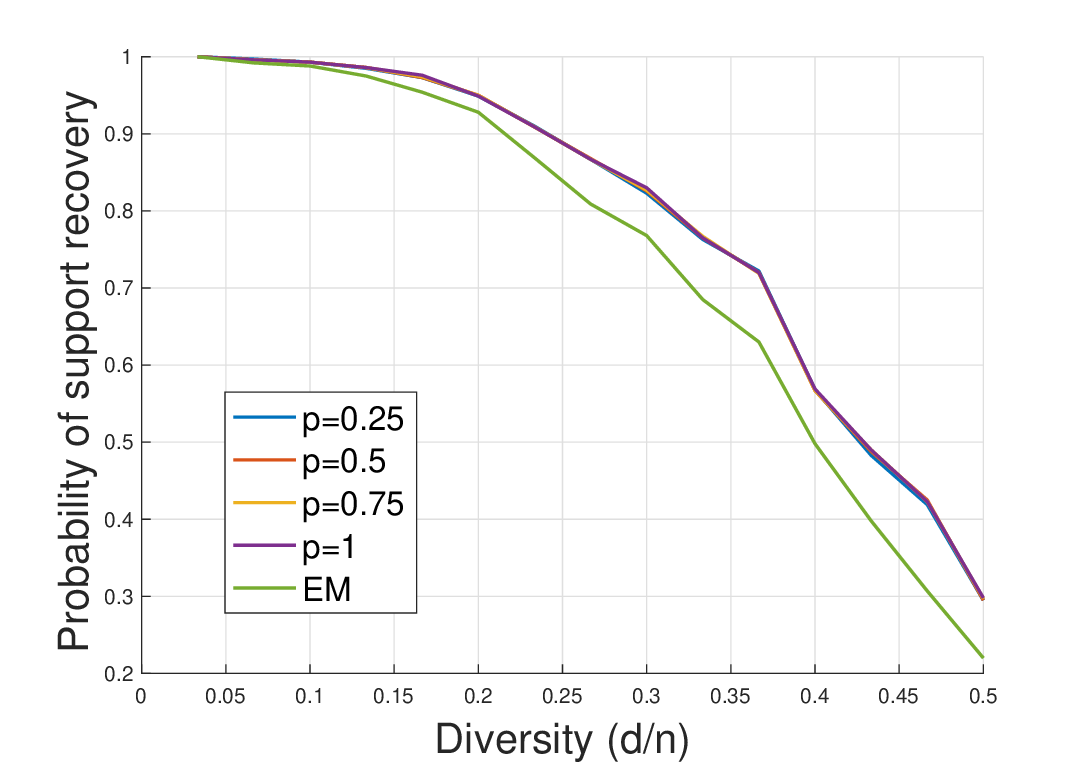}
    \caption{Probability of support recovery}
    \end{subfigure}
    \hspace{1pt}
    \begin{subfigure}[b]{0.4\textwidth}
     \includegraphics[height=4.5cm,width=8cm]{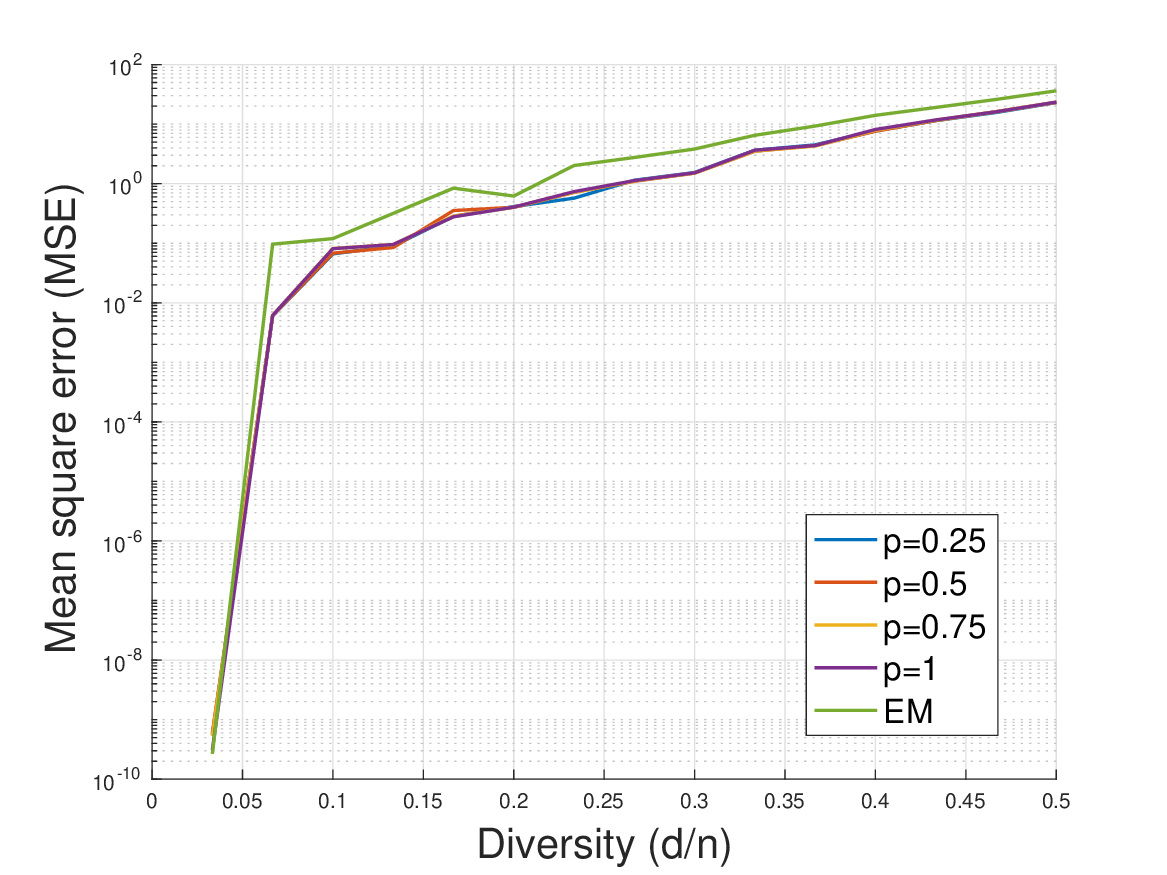}
     \caption{Mean square error}
    \end{subfigure}
    \caption{Performance curves of EM, MU-SBL ($p=1$) and $p$-SBL ($p=\{0.25,0.5,0.75\}$) update rules as a function of the number of non-zero elements for the array matrix ($N$=30, $M$=120, SNR=30 dB).}
    \label{fig:results_classic_alg_arr_mat}
    
\end{figure}

\begin{figure}
    \centering
    \begin{subfigure}[b]{0.4\textwidth}
    \includegraphics[height=4.5cm,width=8cm]{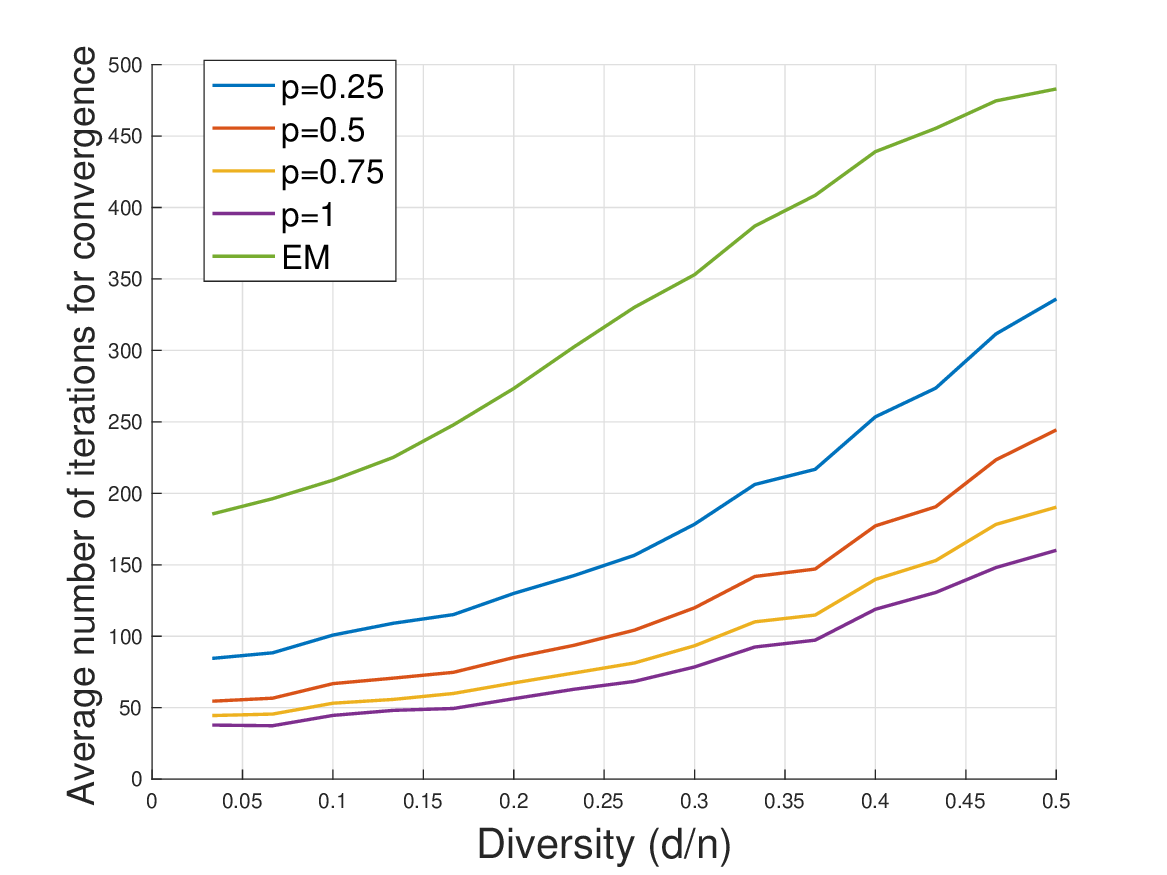}
    \caption{Array Matrix}
    \label{fig:conv_classic_alg_array}
    \end{subfigure}
    \hspace{1pt}
    \begin{subfigure}[b]{0.4\textwidth}
     \includegraphics[height=4.5cm,width=8cm]{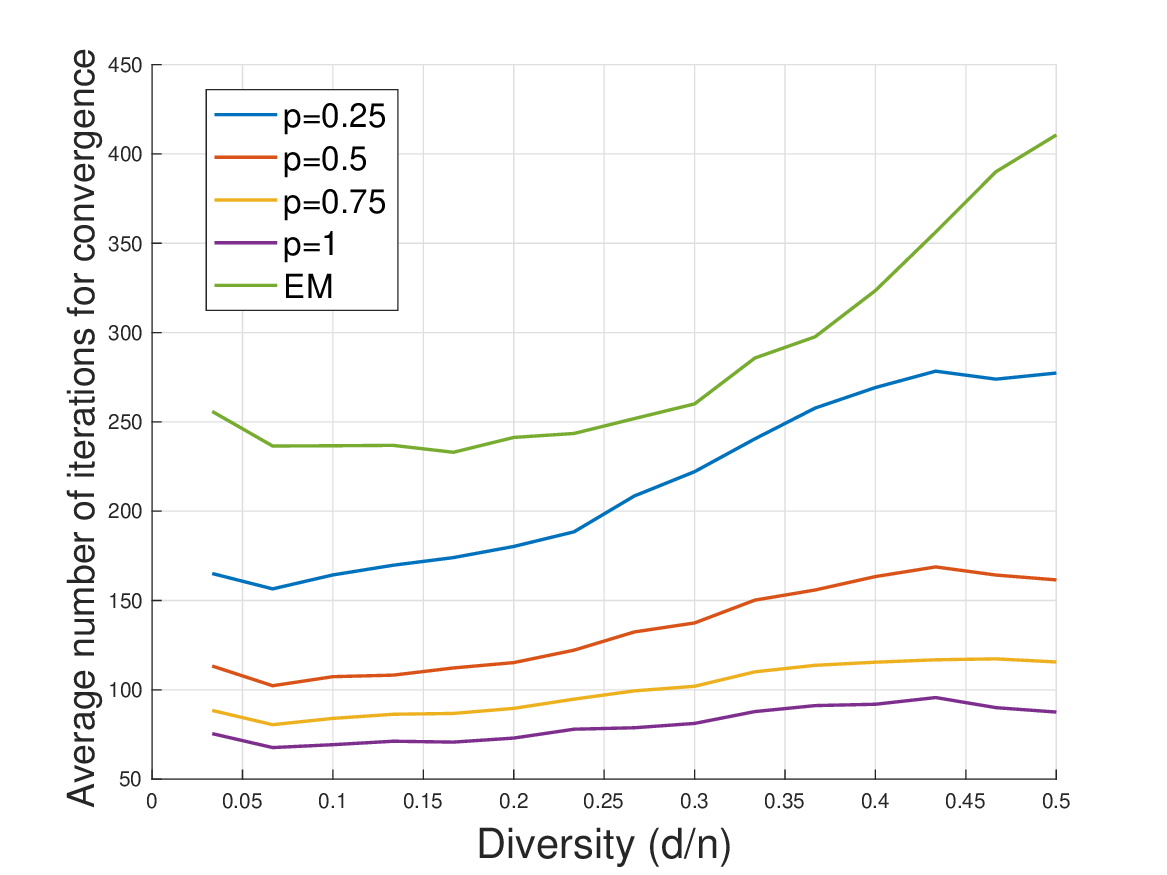}
     \caption{Random matrix}
     \label{fig:conv_classic_alg_rand}
    \end{subfigure}
    \caption{Convergence rates of EM, MU-SBL ($p=1$) and $p$-SBL ($p=\{0.25,0.5,0.75\}$) update rules as a function of the number of non-zero elements ($N$=30, $M$=120, SNR=40 dB).}
    \label{fig:conv_classic_alg}
    
\end{figure}

\begin{figure}
    \centering
\includegraphics[height=5.1cm,width=8cm]{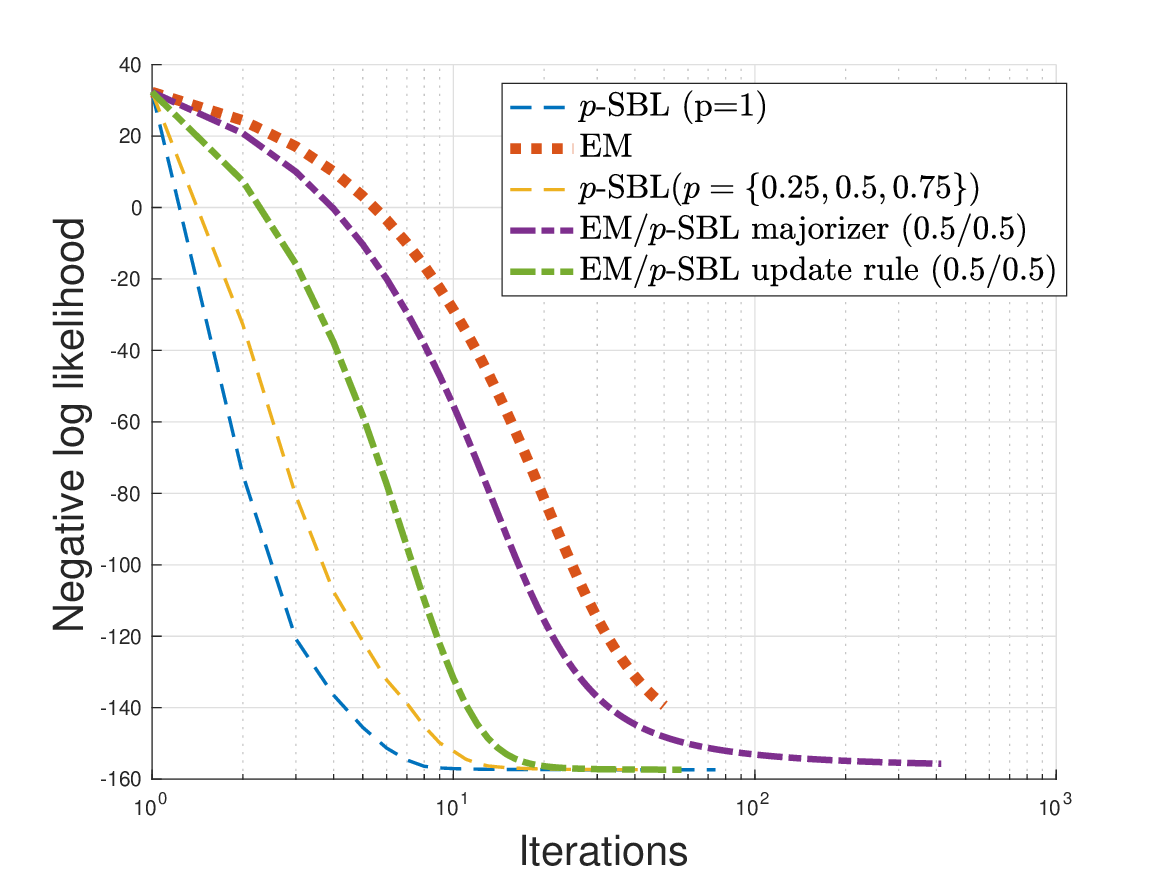}
    \caption{Value of the SBL objective function as a function of the number of iterations for different SBL algorithms for the ULA matrix.}
    \label{fig:nll_convergence}
\end{figure}

{ULA Manifold Matrix}: Fig.~\ref{fig:results_classic_alg_arr_mat} plots the performance curves for EM and the $p$-SBL ($p=\{0.25,0.5,0.75,1\}$) for the array matrix. We see that the $p$-SBL and MU-SBL has the exact same performance and outperforms the EM-SBL. Fig.~\ref{fig:conv_classic_alg_array} plots the average number of iterations  for EM, MU-SBL and $p$-SBL to converge. Even though MU-SBL and $p$-SBL have the exact same performance, the rates of convergence for different values of $p$ are very different. The fastest is the MU-SBL update rule ($p=1$) and the convergence rate decreases as the value of $p$ decreases. The convergence plot shows that there is a mismatch between the curvature of the majorizer and the actual non-convex function $f(\boldsymbol{\gamma})$. According to the majorizer constructed for $p$-SBL, the optimal value of $p$ that minimizes it is $p = 1/2$, and for the case of $p = 1$ the value of the majorizer is unchanged. However, the convergence to a local minimum of $f(\boldsymbol{\gamma})$ as a function of $p$ behaves quite differently. This discrepancy arises because the majorizer's curvature is a local approximation that does not perfectly capture the geometry of the non-convex objective. The $p=1$ update rule produces more aggressive corrections to $\boldsymbol{\gamma}$, which, despite leaving the value of the surrogate unchanged (shown in Appendix~\ref{app:p_sbl}), leads to faster descent on the true objective $f(\boldsymbol{\gamma})$. This observation underscores the practical value of exploring valid descent steps beyond the majorizer's minimizer---a key motivation for the data-driven approach developed in subsequent sections. Fig.~\ref{fig:nll_convergence} shows the value of the SBL ojective function as a function of the number of iterations for different SBL algorithms. As expected from the convergence plots, we see that the $p$-SBL algorithm is the fastest for $p=1$ and EM is the slowest. The convex combination of different $p$-update rules for $p=\{0.25,0.5,0.75\}$ with equal weightings of 1/3 is the fastest after $p=1$. The EM and $p$-SBL update rules ($p=1$) combination with equal weights is the next fastest. The convex combination of the EM and $p$-SBL majorizer 
(equations~(\ref{eq:ema_maj}) and~(\ref{eq:p_sbl_maj})) with equal weighting is the second slowest followed by EM which is the slowest.  

\begin{figure}
    \centering
    \begin{subfigure}[b]{0.4\textwidth}
    \centering
\includegraphics[height=4.5cm,width=8cm]{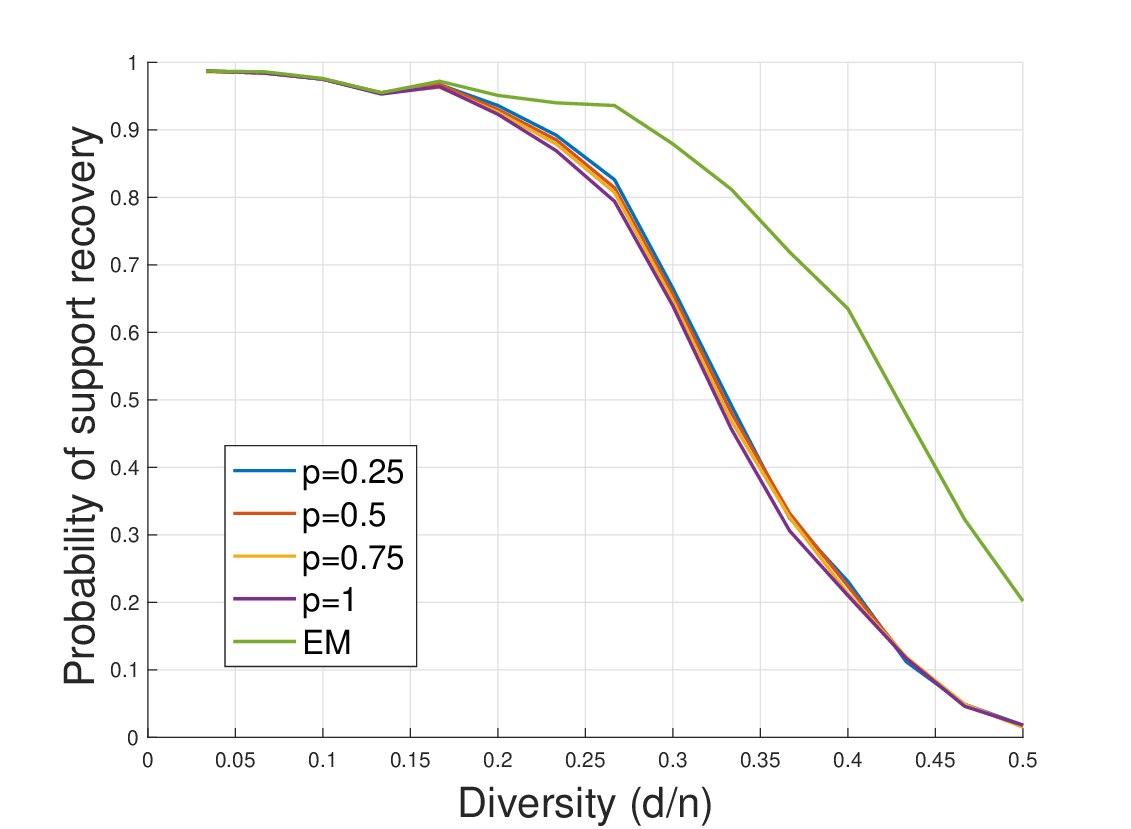}
    \caption{Probability of support recovery}
    \end{subfigure}
    \hspace{3pt}
    \begin{subfigure}[b]{0.4\textwidth}
     \centering\includegraphics[height=4.5cm,width=8cm]{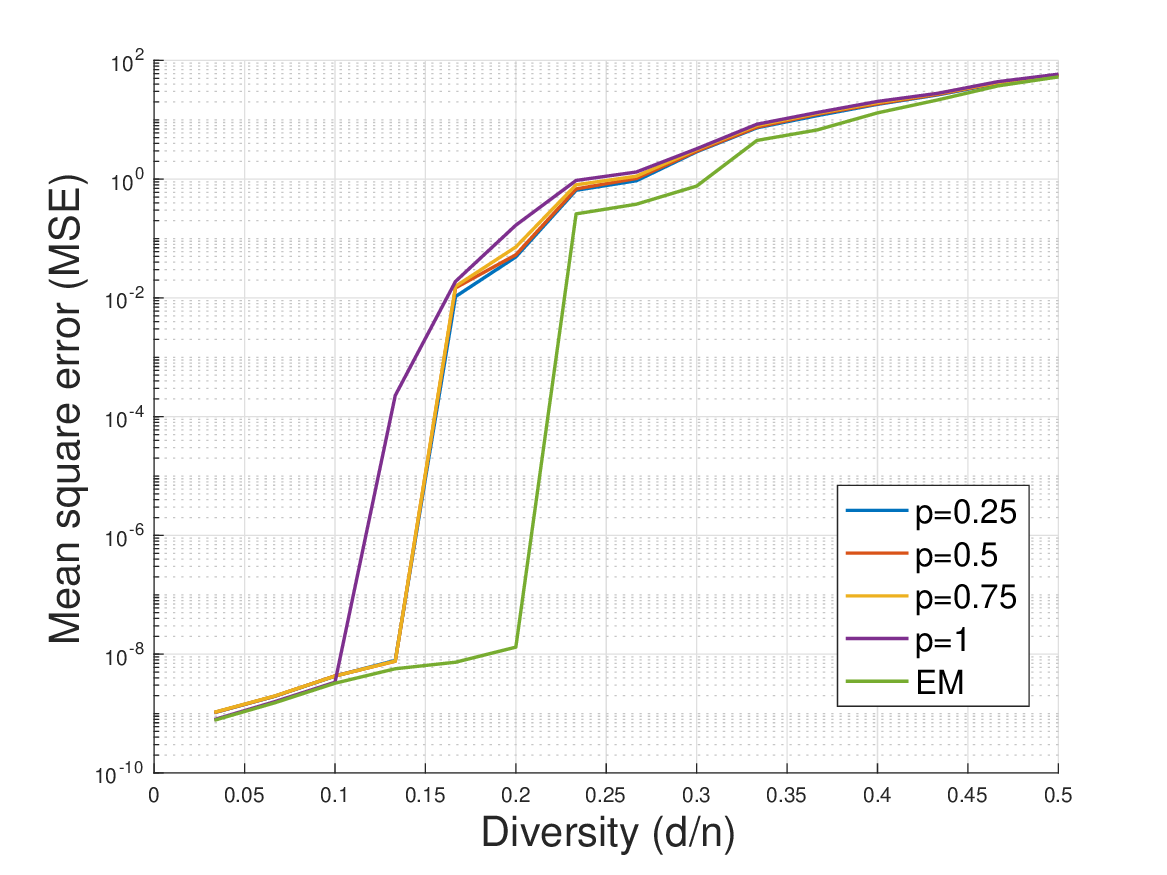}
     \caption{Mean square error}
    \end{subfigure}
    \caption{Performance curves of EM, MU-SBL ($p=1$) and $p$-SBL ($p=\{0.25,0.5,0.75\}$) update rules as a function of the number of non-zero elements for the random matrix($N$=30, $M$=120, SNR=40 dB).}
    \label{fig:results_classic_alg_random_mat}
    
\end{figure}

{Gaussian Random Matrix}: Fig.~\ref{fig:results_classic_alg_random_mat} plots the performance for EM and $p$-SBL ($p\in \{0.25,0.5,0.75,1\}$) for the random matrix case. Unlike the case of the array matrix we see that the EM update rule outperforms the MU-SBL and $p$-SBL. And, for $p$-SBL the performance decreases as $p$ increases. Fig.~\ref{fig:conv_classic_alg_rand} plots the average number of iterations for the random matrix case. Similar to the array matrix case we see that the average number of iterations for convergence increases as $p$ decreases, and the EM algorithm is the slowest. For the case of the random matrix, we see that the EM algorithm on an average takes more iterations to pick out the support set, and it tends to pick the correct one as opposed to the MU-SBL algorithm which converges fast but to the wrong support set more frequently.   

The above results show that the optimal SBL algorithm to choose is a function of the measurement matrix. Hence, it is very difficult to know a-priori, without empirical tests, the best SBL algorithm to choose. To overcome this we choose to learn the best algorithm to pick from data and showcase the results for this in the next section. 
\subsection{Learning the best majorizer via data}
The loss function we are optimizing over to find the best majorizer is the one outlined in~\eqref{eqn:lossfunc}, with $c=0.95$. Adam optimizer~\cite{kingma2014adam} is used to train the model, with a weight decay of $10^{-6}$, learning rate of $4\times 10^{-4}$, and momentum of $0.99$. We first outline the results for the convex combination of the update rules as shown below. 
\subsubsection{Convex combination of the update rules}
\begin{figure*}
\subfloat[Convex combination of update rules from the same majorizer]{\label{fig:weights_convSameMaj}
    \begin{minipage}[b]{0.24\textwidth}
  \centering
  \centerline{\includegraphics[width=0.99\linewidth]{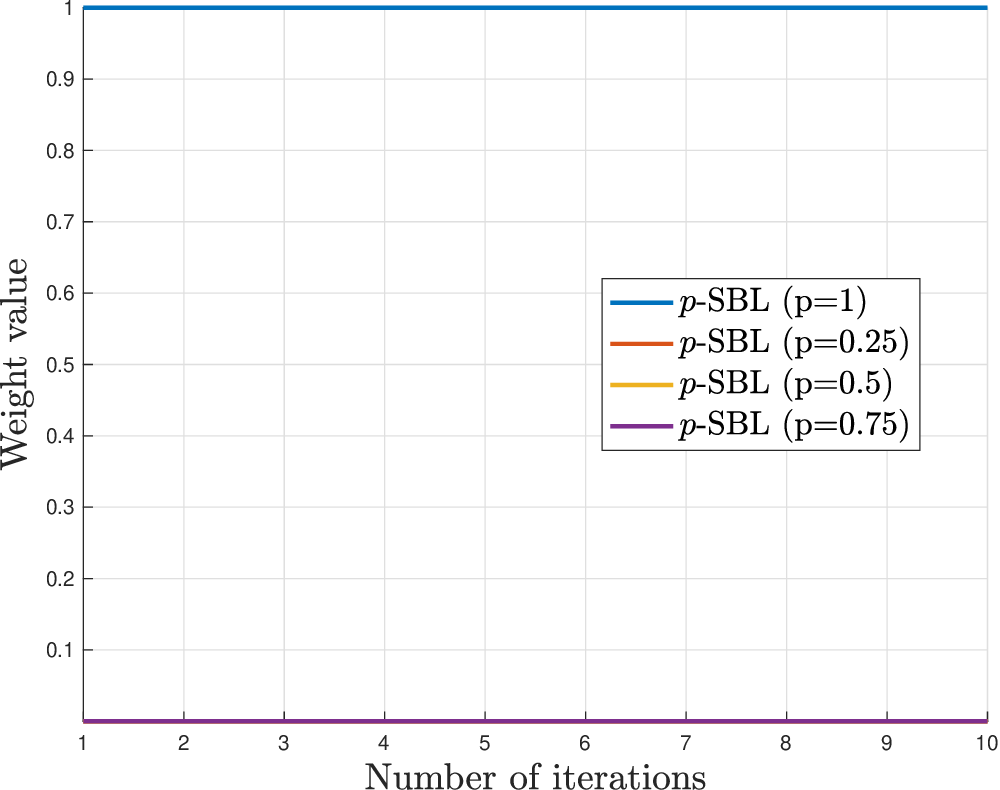}}\smallskip
\end{minipage}
\begin{minipage}[b]{0.24\textwidth}
  \centering
  \centerline{\includegraphics[width=0.99\linewidth]{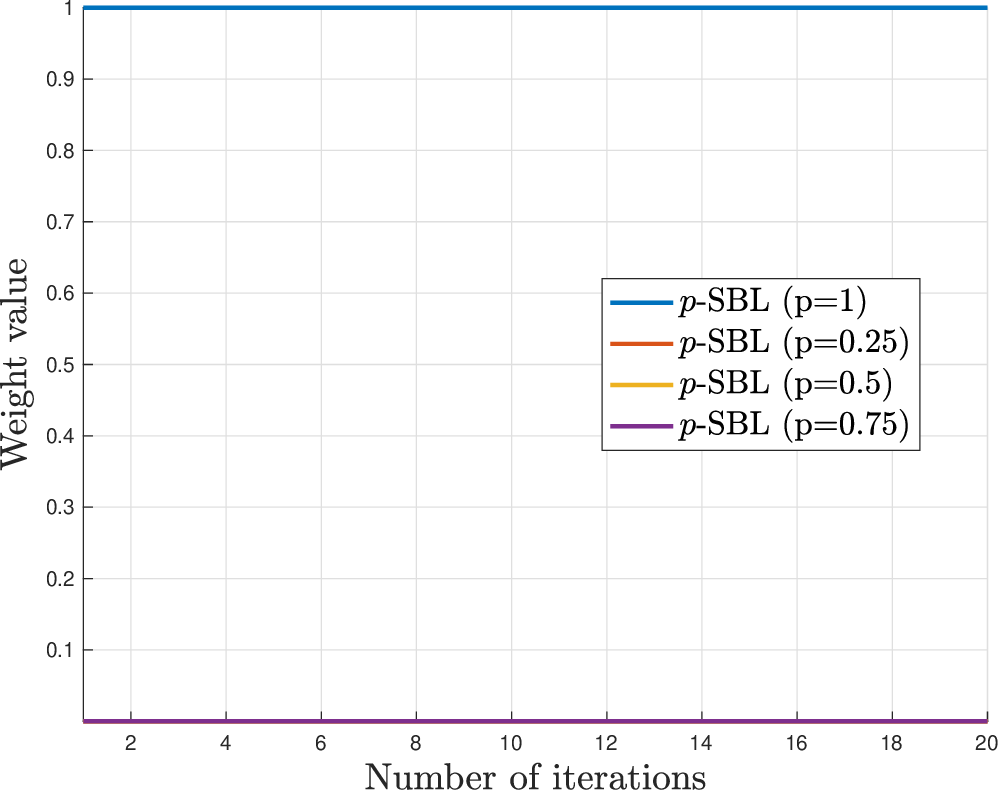}}\smallskip
\end{minipage}

\begin{minipage}[b]{0.24\textwidth}
  \centering
  \centerline{\includegraphics[width=0.99\linewidth]{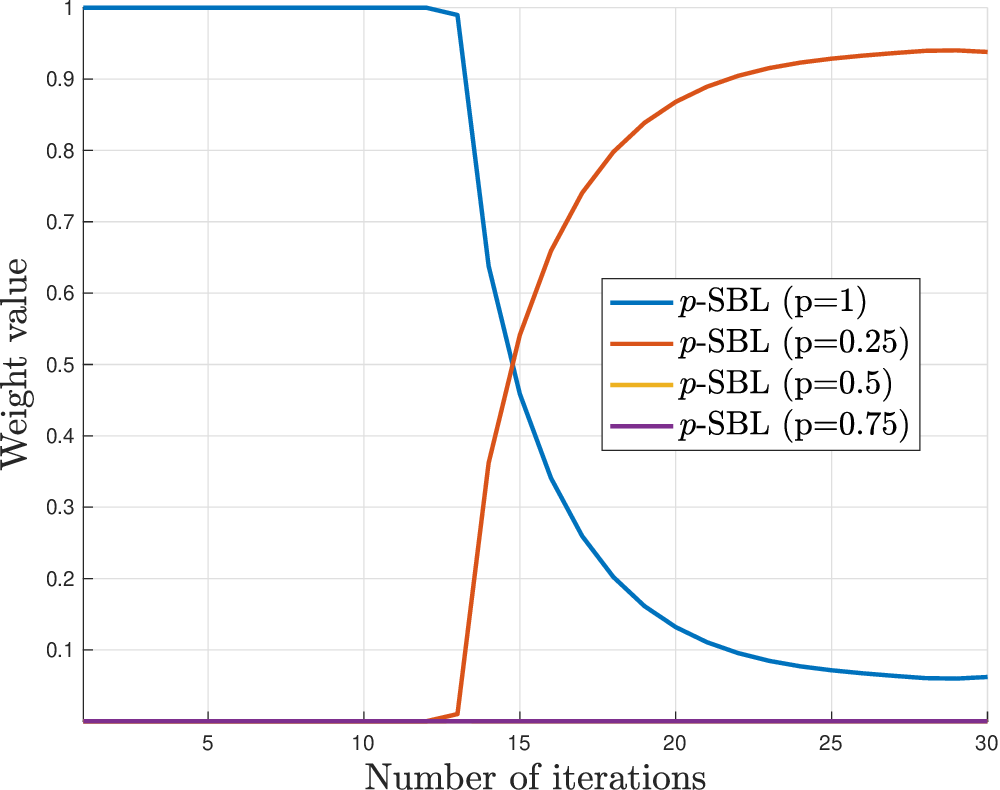}}\smallskip
\end{minipage}
\begin{minipage}[b]{0.24\textwidth}
  \centering
  \centerline{\includegraphics[width=0.99\linewidth]{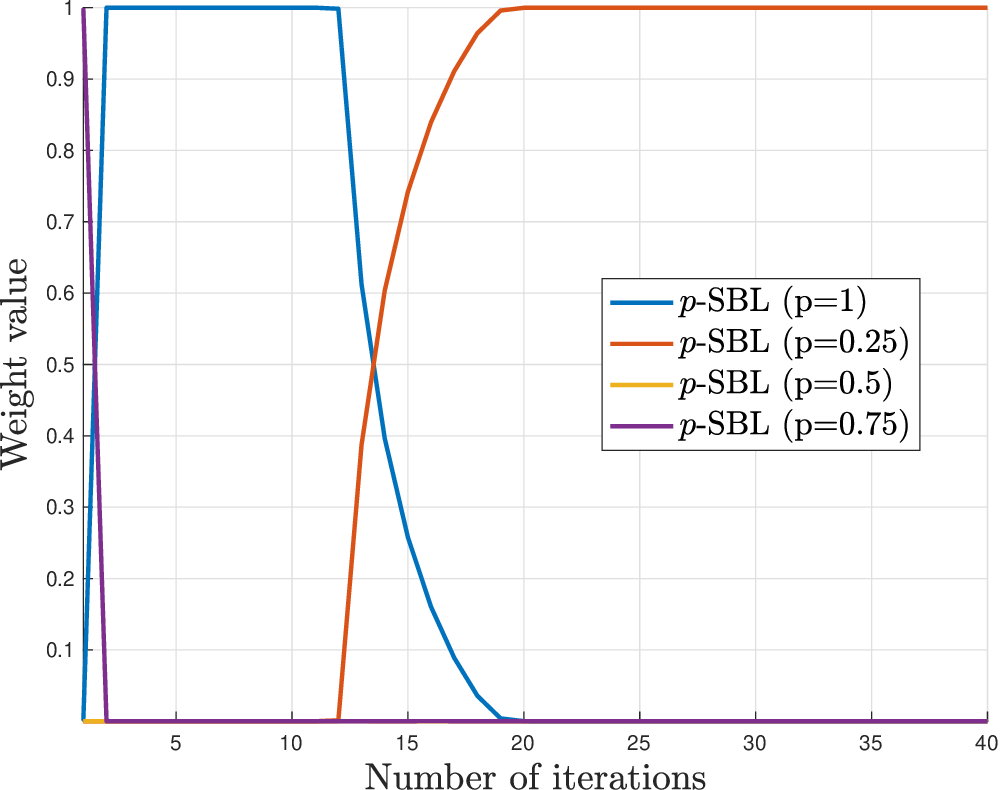}}\smallskip
\end{minipage}}

\subfloat[Convex combination of the EM and $p$-SBL majorizers]{\label{fig:weights_convMaj} \begin{minipage}[b]{0.24\textwidth}
  \centering
  \centerline{\includegraphics[width=0.99\linewidth]{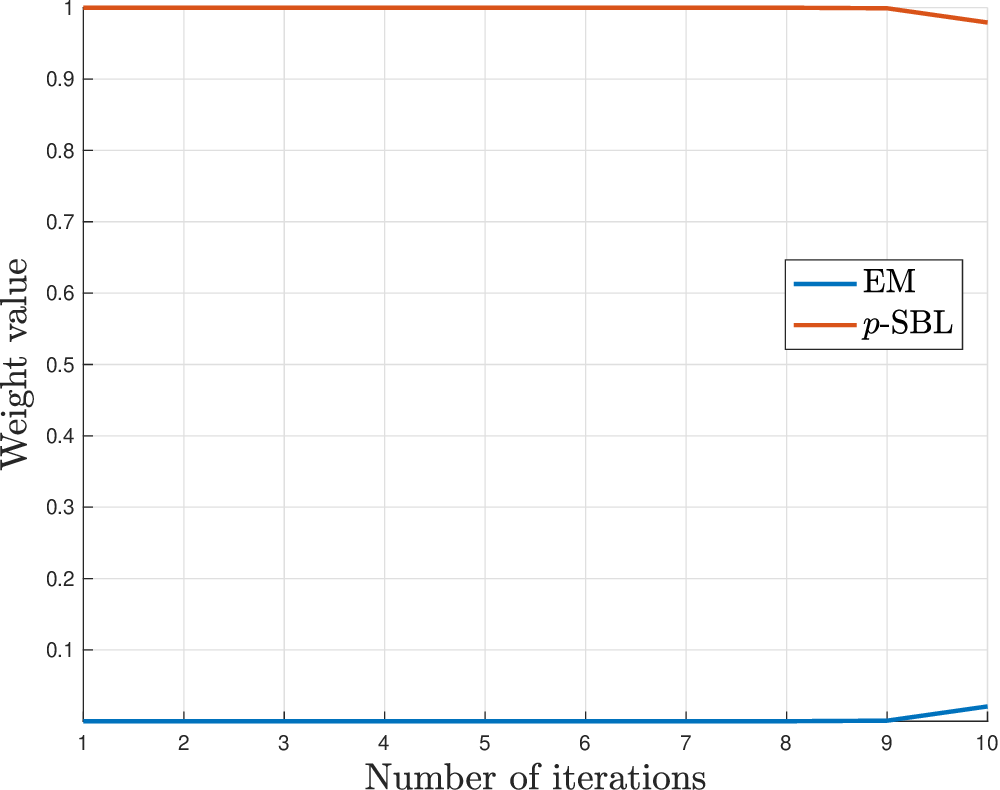}}\smallskip
\end{minipage}
\begin{minipage}[b]{0.24\textwidth}
  \centering
  \centerline{\includegraphics[width=0.99\linewidth]{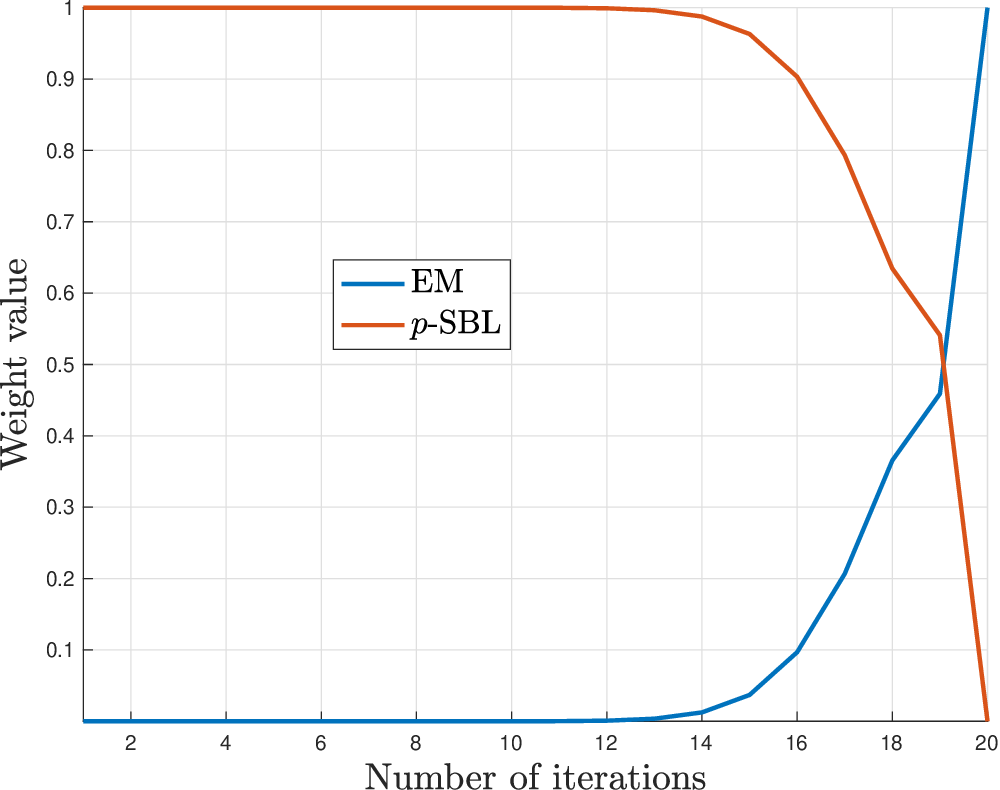}}\smallskip
\end{minipage}

\begin{minipage}[b]{0.24\textwidth}
  \centering
  \centerline{\includegraphics[width=0.99\linewidth]{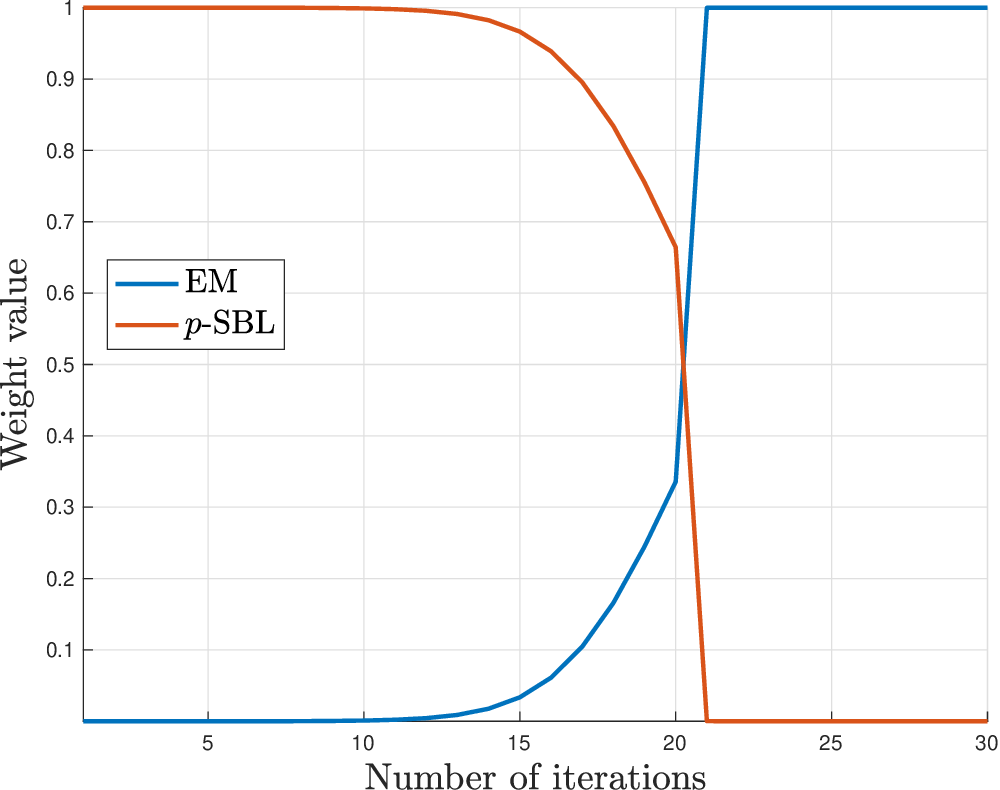}}\smallskip
\end{minipage}
\begin{minipage}[b]{0.24\textwidth}
  \centering
  \centerline{\includegraphics[width=0.99\linewidth]{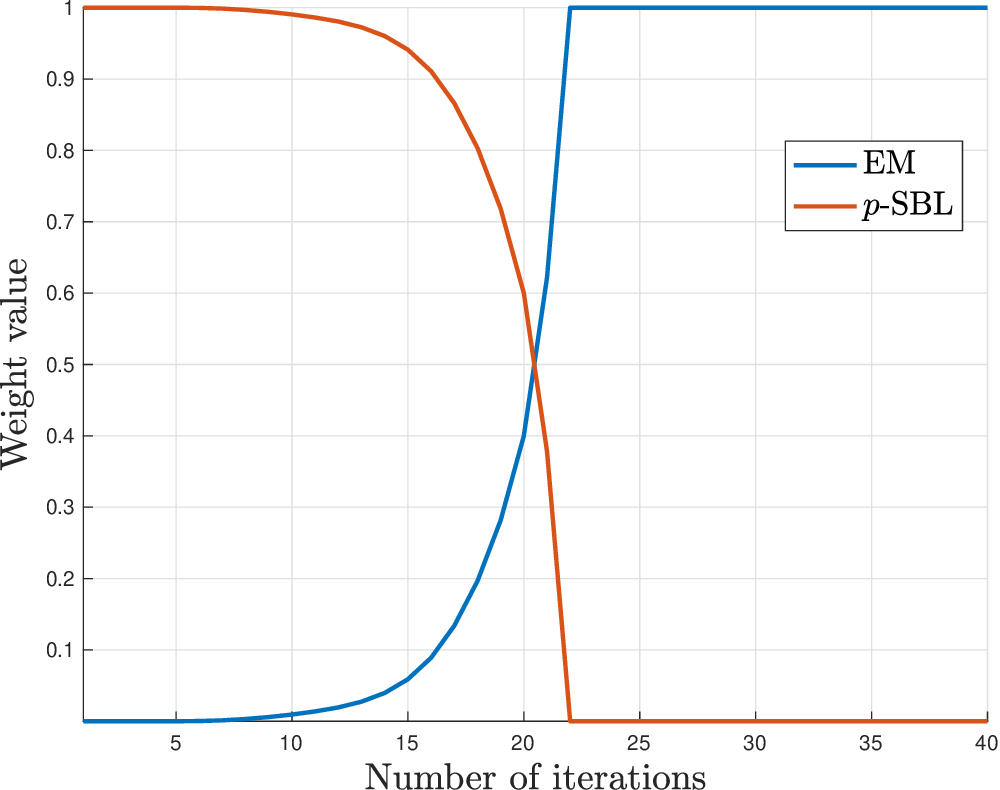}}\smallskip
\end{minipage}
}
\caption{Evolution of the weights as a function of iterations. We showcase the results for different number of iterations as well as the different types of update rule generation for the array matrix ($N=30,M=120$, SNR=40~dB).} 
\label{fig:weight_evolution}
\end{figure*}

In this section we outline the two different cases considered, first we combine update rules corresponding to the same majorizer and in the next case we consider combining update rules from different majorizers. As an example of the former, we combine the update rules corresponding to $p$-SBL for different values of $p$. We choose $p=\{0.25,0.5,0.75,1\}$. We fix the number of MM iteration to a constant $J$. The $j^{\text{th}}$ iteration has the following non-negative learnable weights $\{\alpha^i_{\text{0.25}},\alpha^i_{\text{0.5}},\alpha^i_{\text{0.75}},\alpha^i_{\text{1}}\}$, all of which add up to 1. Figure~\ref{fig:weights_convSameMaj} shows the evolution of the weights for $J=\{10,20,30,40\}$. When the number of iterations is low, the network only uses the update rule that converges the fastest ($p=1$), however upon increasing the number of iterations ($J\in\{20,40\}$) we see that the network chooses the fastest update rule in the beginning but in the last few iterations it switches to the slowest update rule ($p=0.25$).

\subsubsection{Convex combination of majorizers}
Here we combine the majorizers for EM-SBL and $p$-SBL (\eqref{eq:ema_maj} and~\eqref{eq:p_sbl_maj}).
We unroll~\eqref{eq:convMaj_updateRule} and fix the number of iterations of the MM algorithm to $J$. We have a total of $J \times 2$ parameters, where the weights at each iteration are constrained to be non-negative and sum up to 1. Fig.~\ref{fig:weights_convMaj} show the weights as a function of iterations for different number of maximum iterations. We see similar behavior in terms of the choice of the majorizer, when the iteration number is fixed to 10, the fastest majorizer is chosen but as the maximum iterations is increased we see that the fastest majorizer is chosen in the initial few iterations and in the last few, the algorithm switches to the slowest one.     
\subsection{DNN-SBL with Skip Connection}
In this section we showcase the results of our DNN-SBL model. The number of iterations is fixed to $J$=15. In each iteration, the dimension $d$ is 64. The exponential constant used in the $w$-MSE loss is $c=$0.95. For the skip connection we consider a general update rule by combining the EM update rule and $p$-SBL update rules for $p=\{0.25,0.5,0.75,1\}$ using the weights $\{\alpha_1,\ldots,\alpha_5\}$,
where $\alpha_i\geq 0\;\forall\;i\in\{1,\ldots,5\}$ and $\sum^5_{i=1}\alpha_i=1$. We learn the $\alpha$'s along with the weights of the neural network using the strategy illustrated in section~\ref{sec:conv_maj}. 

\begin{figure*}
\subfloat[Array Matrix ($N$=30, $M$=120, $\text{SNR}$=30 dB)]{\label{fig:array}
    \begin{minipage}[b]{0.27\textwidth}
  \centering
  \centerline{\includegraphics[width=0.99\linewidth]{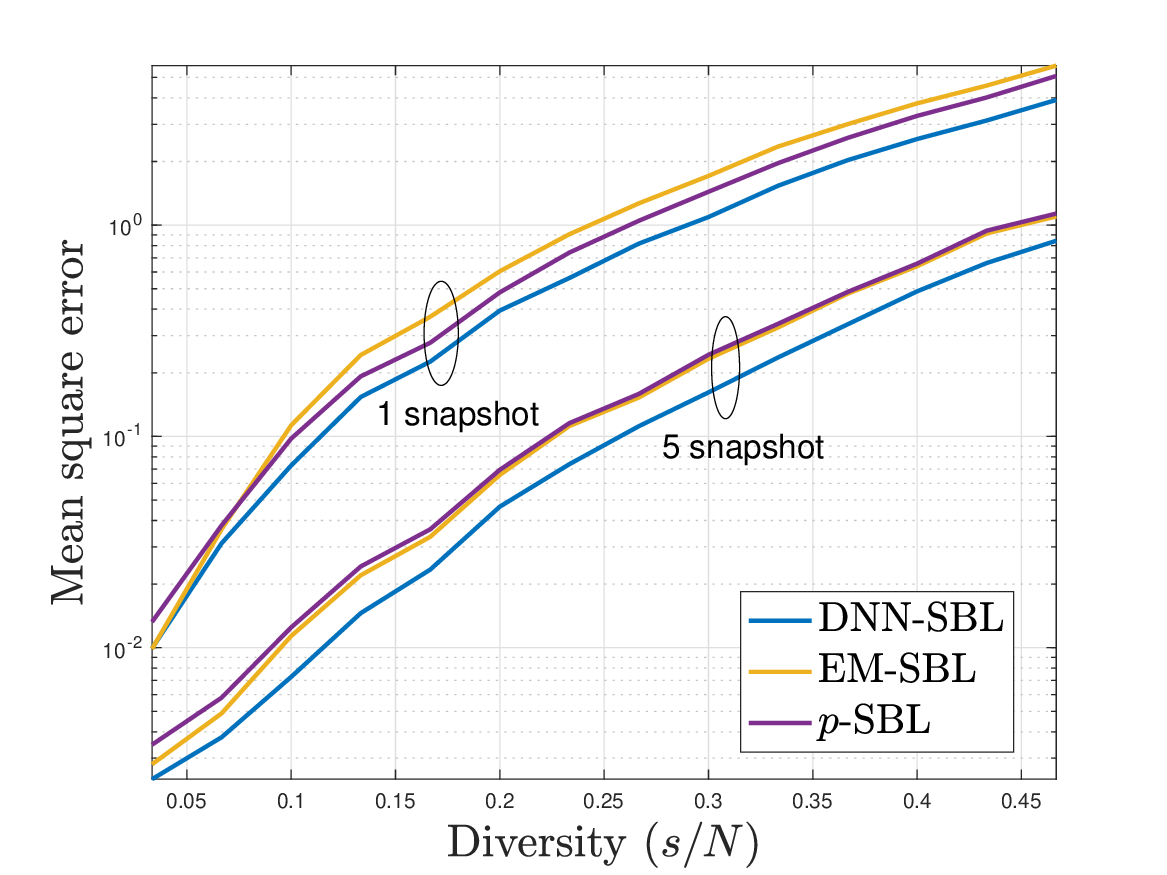}}\smallskip
\end{minipage}
\hspace{-20pt}
\begin{minipage}[b]{0.27\textwidth}
  \centering
  \centerline{\includegraphics[width=0.99\linewidth]{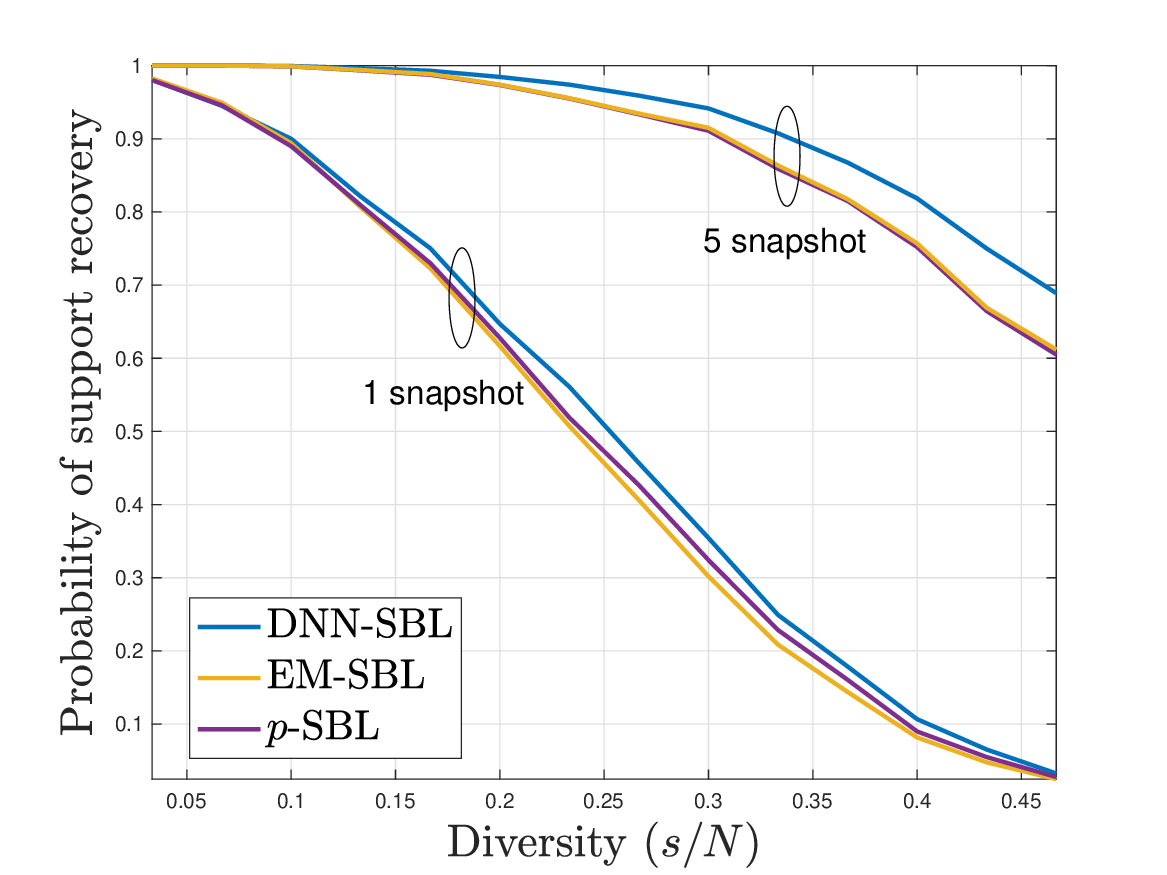}}\smallskip
\end{minipage}
}
\hspace{-20pt}
\subfloat[Complex random matrix($N$=30,$M$=120,$L$=1)]{\label{fig:c_rand}
\begin{minipage}[b]{0.27\textwidth}
  \centering
  \centerline{\includegraphics[width=0.99\linewidth]{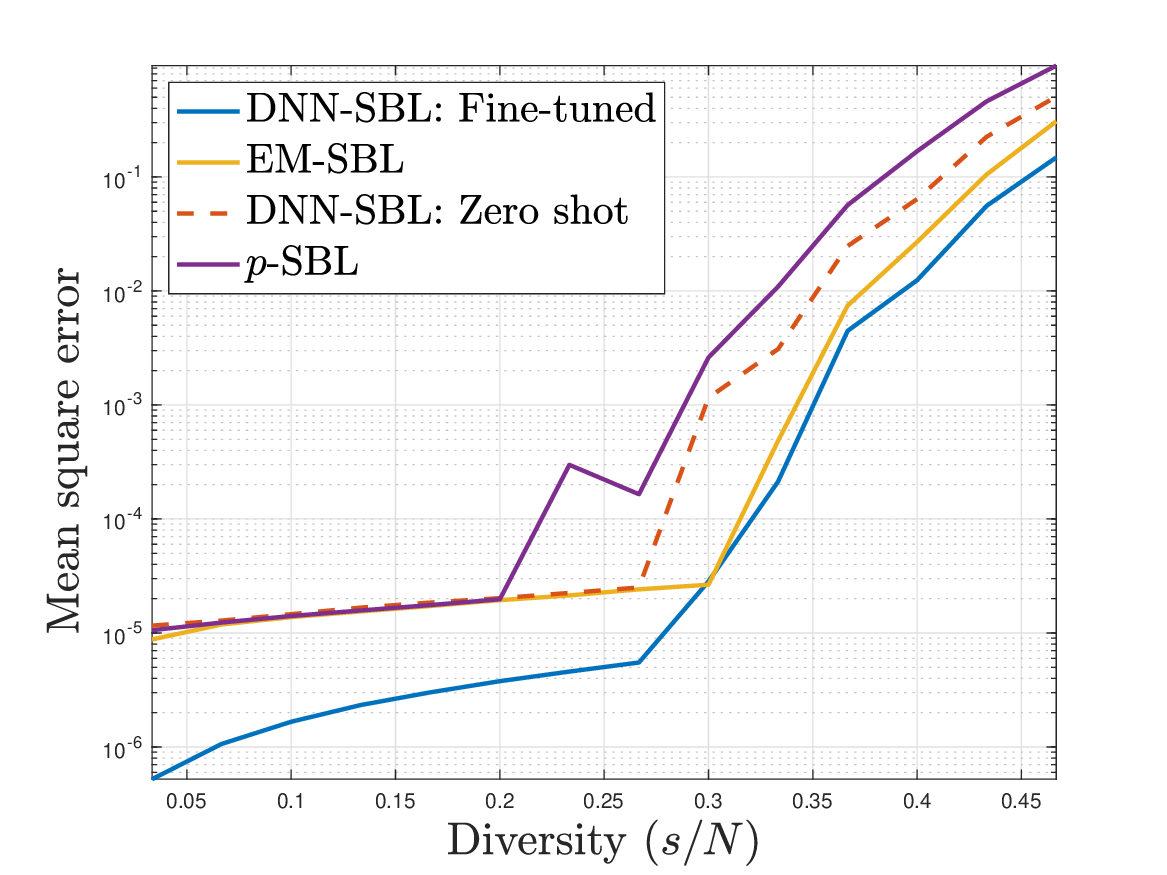}}\smallskip
\end{minipage}
\hspace{-20pt}
\begin{minipage}[b]{0.27\textwidth}
  \centering
  \centerline{\includegraphics[width=0.99\linewidth]{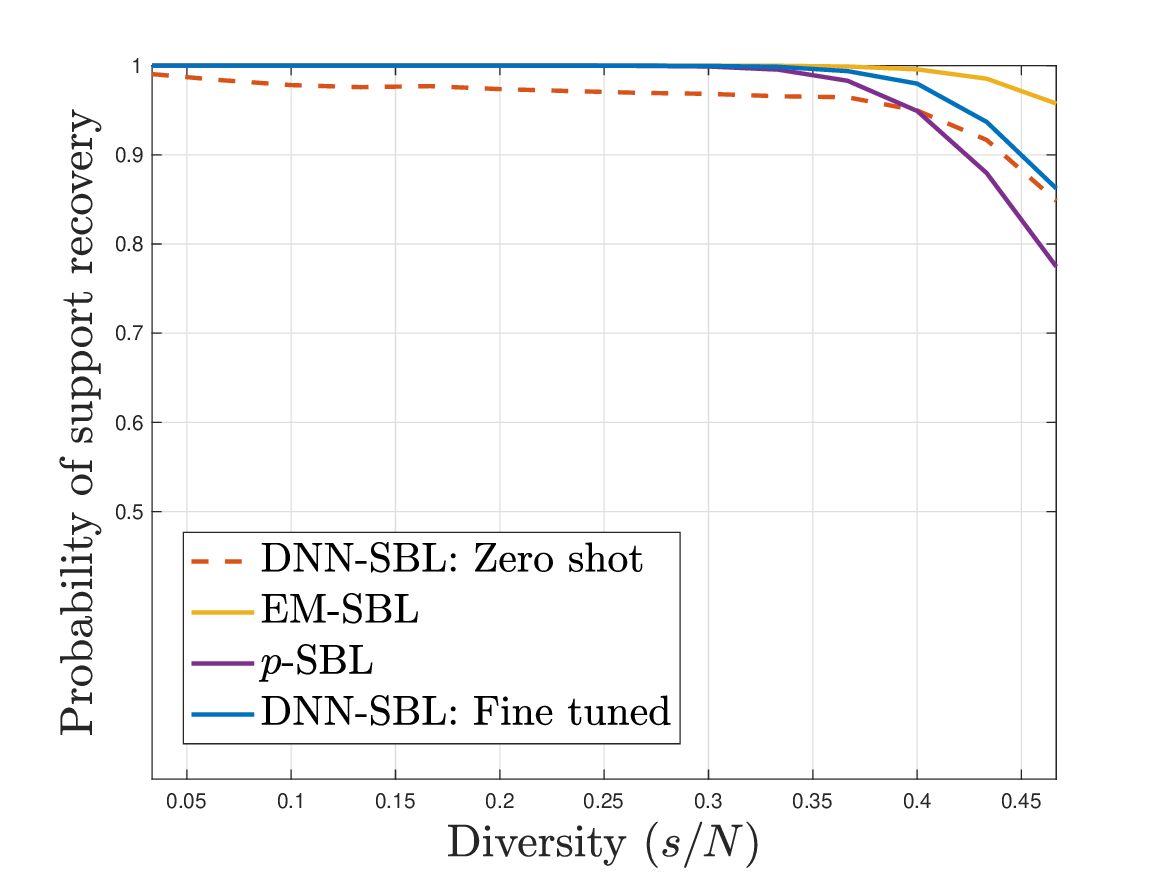}}\smallskip
\end{minipage}
}


\subfloat[Correlated matrix ($N$=20, $M$=100, $L$=7)]{\label{fig:corr} \begin{minipage}[b]{0.27\textwidth}
  \centering
  \centerline{\includegraphics[width=0.99\linewidth]{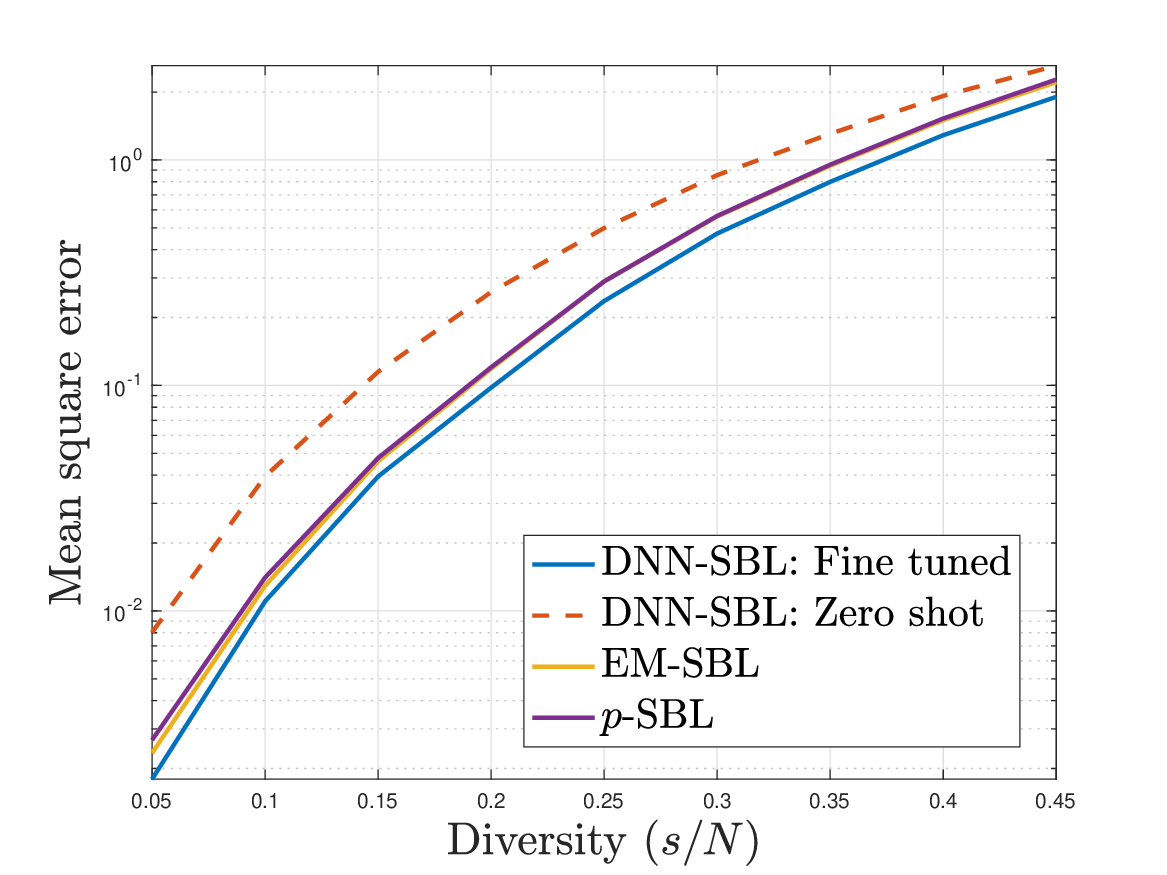}}\smallskip
\end{minipage}
\hspace{-20pt}
\begin{minipage}[b]{0.27\textwidth}
  \centering
  \centerline{\includegraphics[width=0.99\linewidth]{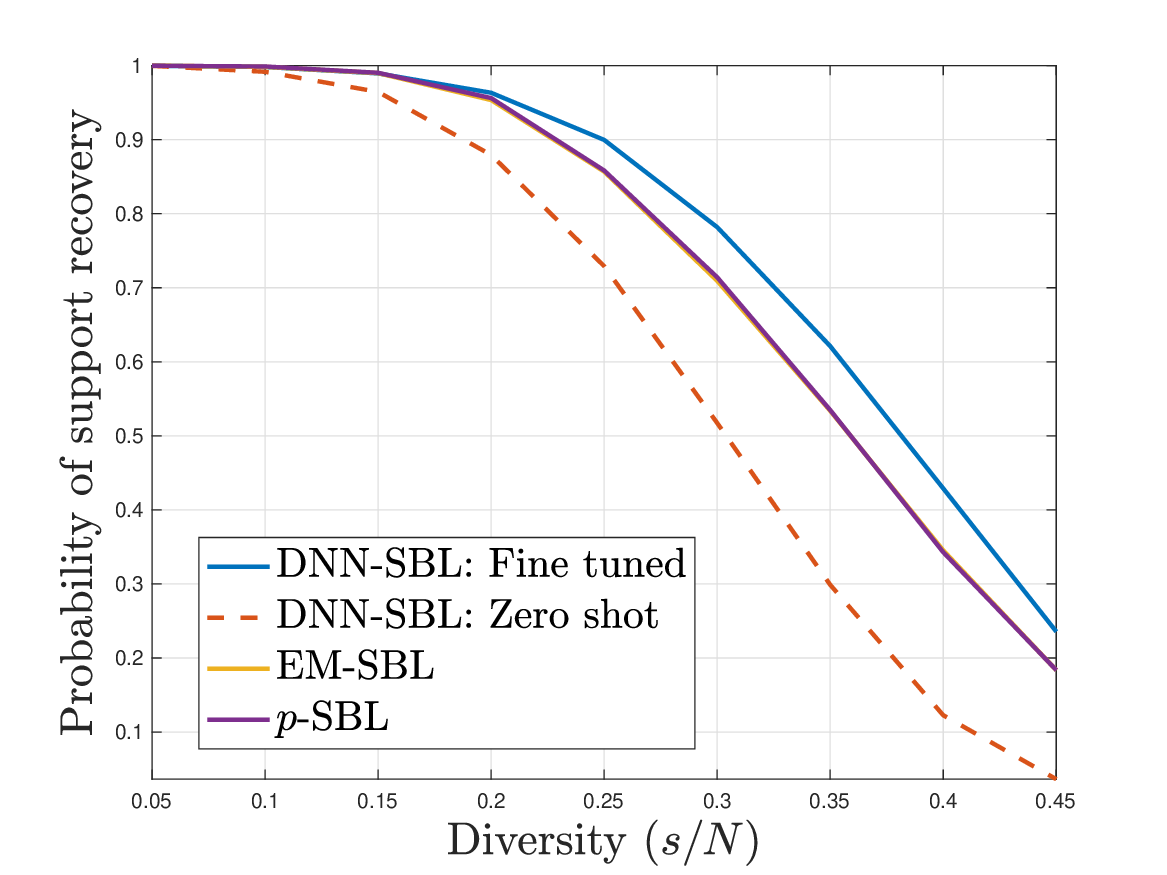}}\smallskip
\end{minipage}
}
\hspace{-20pt}
\subfloat[Unseen array matrix ($N$=30, $M$=180, $L$=3)]{\label{fig:array180}\begin{minipage}[b]{0.27\textwidth}
  \centering
  \centerline{\includegraphics[width=0.99\linewidth]{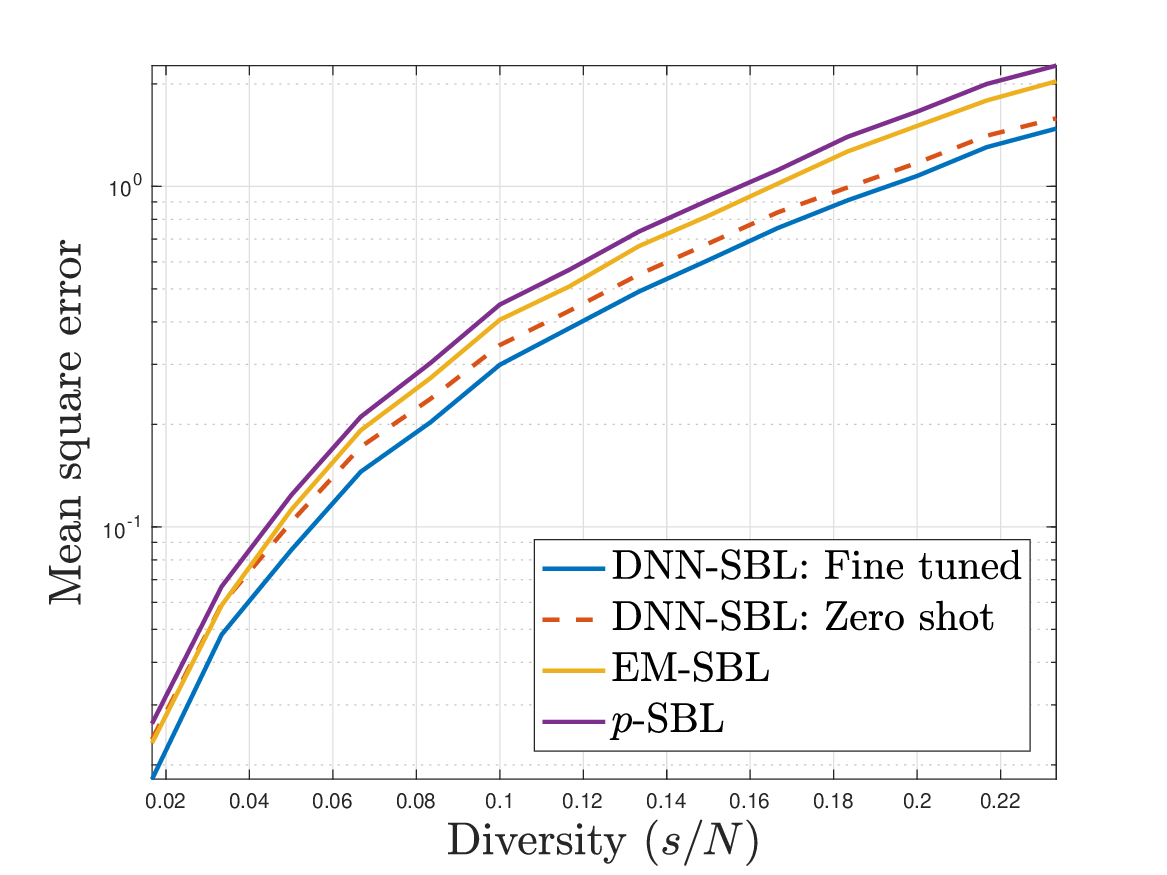}}\smallskip
\end{minipage}
\hspace{-20pt}
\begin{minipage}[b]{0.27\textwidth}
  \centering
  \centerline{\includegraphics[width=0.99\linewidth]{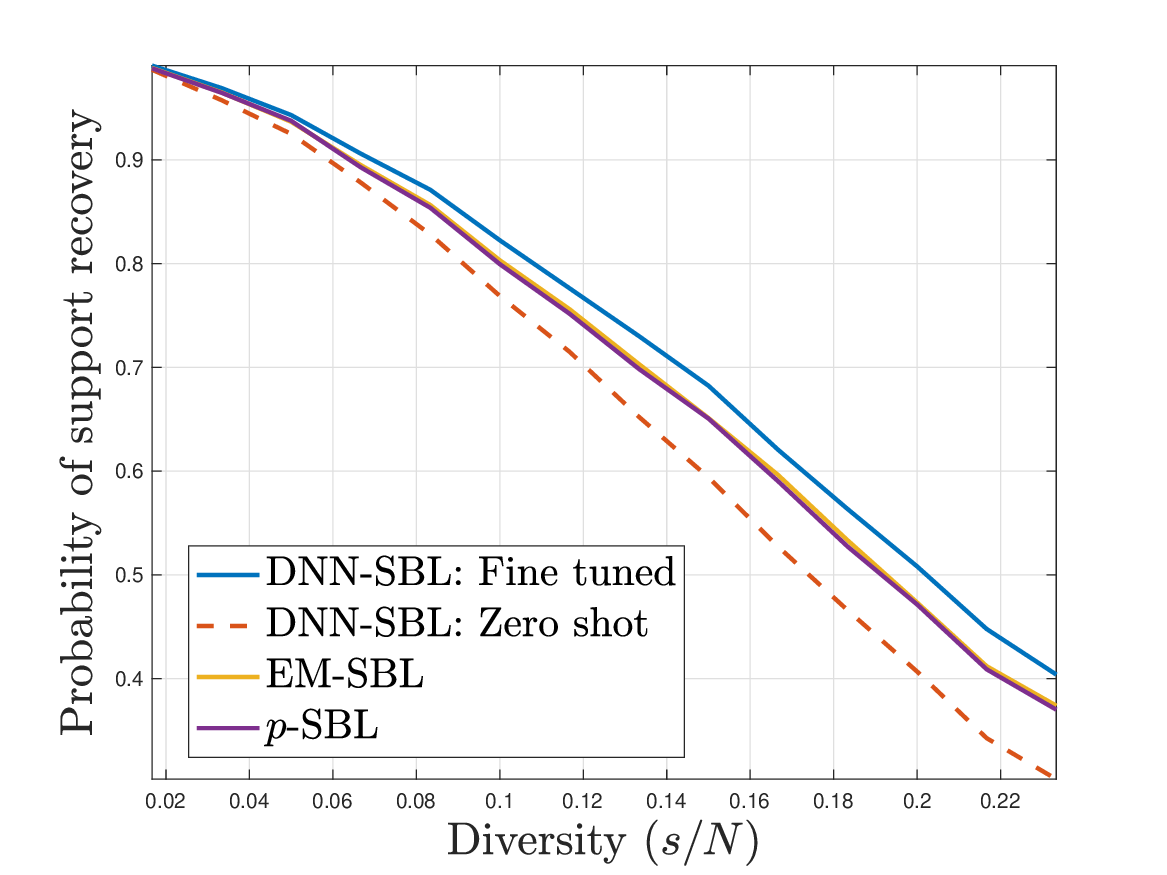}}\smallskip
\end{minipage}
}
\caption{Performance of our proposed neural network architecture on different types of measurement matrices.} 
\label{fig:perf_nn}
\end{figure*}
\paragraph{Training} We train the network on a dataset of size $1.4\text{M}$ samples generated using the array matrix of size $N=30$, $M=120$, $L=[1,2,5,7,10]$ and $\text{SNR}=30$~dB, where each snapshot has the same number of samples. The grid boundaries are given by the parameters $\beta_1=31^o$ and $\beta_{120}=150^o$ with a grid spacing of $1^{o}$. The network is trained for 200 epochs using the AdamW optimizer~\cite{loshchilovdecoupled} and cosine learning rate scheduler with a learning rate of $2\times 10^{-4}$ with the co-sine annealing scheduler, weight decay is not used and the batch size is 2048.  

\paragraph{Testing}We test the network across different diversity levels (number of non-zero elements in $\boldsymbol{\gamma_{*}}$) using the MSE and PSR as the metrics and the results are shown in Fig.~\ref{fig:array}. \newline\textit{Superior update rule:} We see that our network named DNN-SBL outperforms the popular SBL algorithms across different diversity levels for both the single and multi-snapshot case. \newline\textit{Measurement matrix generalization:} We demonstrate the generalization to unseen measurement matrices by testing the network trained on $\boldsymbol{\Phi}_{A}$ but tested on $\boldsymbol{\Phi}_R$,  and $\boldsymbol{\Phi}_C$. Fig.~\ref{fig:c_rand} shows the zero-shot performance of DNN-SBL on the complex random matrix for $\text{SNR}=60$~dB, $L=1$, $N=30$ and $M=120$. We see that the network performs well when presented with an unseen matrix and fine-tuning it on a small dataset generated using $\boldsymbol{\Phi}_R$ outperforms the popular SBL algorithms in terms of MSE and matches the EM-SBL in terms of PSR except in the high-sparsity regime. \newline\textit{Modular design:} To showcase the flexibility of our network we test the network on the correlated matrix of size $N=20,M=100$ with $\text{SNR}=40$~dB and $L=7$. No architectural changes are required to test the model in the above scenario and the results are shown in~Fig.~\ref{fig:corr}. The zero-shot performance in terms of PSR is close to the popular algorithms in the low diversity regime but on fine-tuning the model we achieve superior performance over all diversity levels. An advantage of this flexibility arises in the array signal processing domain, where the model trained on a region between $31^o$ to $150^o$ can now be tested on a grid spanning $0^o$ to $180^o$ with a grid spacing of $1^o$. We show the results of such an experiment in Fig.~\ref{fig:array180} where the $\text{SNR}=30$~dB and $L=3$. The model performs better than the popular SBL algorithms in terms of MSE but is slightly worse off in PSR, however the fine-tuned network achieves the best performance in both the metrics. We see similar behaviour in terms of zero-shot performance when the base model is trained on the correlated matrix and tested on the different types of sensing matrices.

\section{Conclusion and Future Work}
In this work we developed a novel re-parameterization of the SBL algorithm, and provided a beamforming interpretation to the update rules corresponding to the EM-SBL and MU-SBL. Using the MM framework, we introduced a new class of SBL algorithms called $p$-SBL. This expanded class contained the MU-SBL update rule, hence providing convergence guarantees to it. The $p$-SBL majorizer not only brings the two most popular SBL algorithms under the MM framework but the two update rules correspond to minimizing the same majorizer. Additionally, using MM theory we showed how one could expand the class of existing majorizers and update rules. Furthermore, using data-driven techniques we learned an SBL algorithm from this expanded class of majorizers. We then go beyond the class of majorizers to learn a more general update rule by parameterizing the class with a  novel neural network architecture. We show that the network has many interesting attributes; complexity of the architecture is invariant to the size of the measurement matrix and the number of snapshots. Finally, we showed that the network trained on a parameterized dictionary can not only generalize to the same dictionary type of a different size, but also generalize to unseen measurement matrices. The proposed DNN-SBL architecture represents an important first step toward learning SBL algorithms. The simplicity gained by the $\mathbf{T}_1$ and $\mathbf{T}_2$ parameterization, which enabled the dimensional invariance of our architecture, can also be a hindrance in practical scenarios where the true measurement matrix is unavailable and only an idealized or approximate version of it is accessible. In such situations, $\mathbf{T}_1$ and $\mathbf{T}_2$, being computed from the assumed matrix, may not capture the additional variability introduced by the mismatch, potentially preventing the network from compensating for it.  Studying the robustness of the proposed framework under measurement matrix uncertainties and developing architectures that can account for such mismatches remains an important direction for future work. We expect that with application-specific datasets and tailored architectural choices, further performance gains can be expected.


%

\appendices

\vspace{-0.2cm}
\section{Proof of $p$-SBL algorithm}
\label{app:p_sbl}

For the sake of brevity we choose the case when $L=1$ to illustrate the proof, however the proof can be easily extended to the case when $L>1$. As derived in~\cite{wipf}, the data dependent in $f(\boldsymbol{\gamma})$ can be obtained as a solution to the following optimization problem
    \begin{equation}
    \begin{split}
        \vecsym{y}^{H}\covyy^{-1}(\boldsymbol{\gamma})\vecsym{y} &= \min_{\vecsym{x}} \frac{1}{\sigma^2}||\vecsym{y}-\mm \vecsym{x}||^2_2 + \sum^{M}_{i=1} \frac{\mathbf{x}^2[i]}{\boldsymbol{\gamma}[i]}\\
        &\leq  \frac{1}{\sigma^2}||\vecsym{y}-\mm \hat{\vecsym{x}}_j||^2_2 + \sum^{M}_{i=1} \frac{\hat{\mathbf{x}}_j^2[i]}{\boldsymbol{\gamma}[i]}
        \end{split}
    \end{equation}
    where $\hat{\vecsym{x}}_j=\boldsymbol{\hat{\Gamma}}_j\mm^{H}\covyy^{-1}(\boldsymbol{\hat{\gamma}}_j)\vecsym{y}$, is the conditional mean. The log-det term in $f(\boldsymbol{\gamma})$ is a concave function and can be bounded by a supporting hyperplane at any point $\boldsymbol{\gamma}_j$.
    \begin{equation}
        \label{eq:logdet_mm}
        \log|\mm \mathbf{\Gamma} \mm^{H} + \sigma^2\mathbf{I}| \leq \sum^{M}_{i=1}\boldsymbol{\gamma}[i] \mathbf{T}_2(\boldsymbol{\hat{\gamma}}_j)[i] + \mbox{const}
    \end{equation}
    Combining the above two equations, simplifying, and retaining terms involving $\boldsymbol{\gamma}$, we obtain 
    the upper bound as
     \begin{equation}
         f(\boldsymbol{\gamma})\leq g(\boldsymbol{\gamma};\boldsymbol{\hat{\gamma}}_j) = \sum^{M}_{i=1} g_i(\boldsymbol{\gamma}[i];\boldsymbol{\hat{\gamma}}_j) + c,
         \end{equation}
        where $g_i(\boldsymbol{\gamma}[i];\boldsymbol{\hat{\gamma}}_j) = \boldsymbol{\gamma}[i] \mathbf{T}_2(\boldsymbol{\hat{\gamma}}_j)[i] + \frac{\boldsymbol{\hat{\gamma}}^2_j[i]\mathbf{T}_1(\boldsymbol{\hat{\gamma}}_j)[i]}{\boldsymbol{\gamma}[i]}$,
     and $f(\boldsymbol{\hat{\gamma}}_j)=g(\boldsymbol{\hat{\gamma}}_j;\boldsymbol{\hat{\gamma}}_j)$. 

     Now we show that for the general $p$-SBL update rule given in (\ref{eq:p_t_update}), $\boldsymbol{\hat{\gamma}}_{j+1} = \left(\frac{\mathbf{T}_1(\boldsymbol{\hat{\gamma}}_j)}{\mathbf{T}_2(\boldsymbol{\hat{\gamma}}_j)}\right)^p \boldsymbol{\hat{\gamma}}_j,\; 0 < p \leq 1,$
     we have descent, if $\boldsymbol{\hat{\gamma}}_j$ is not a a stationary point ($\mathbf{T}_2(\boldsymbol{\hat{\gamma}}_j) \neq  \mathbf{T}_1(\boldsymbol{\hat{\gamma}}_j)$), i.e. 
     \begin{equation}\label{eq:mm_cond}
g(\boldsymbol{\hat{\gamma}}_{j+1};\boldsymbol{\hat{\gamma}}_j) < g(\boldsymbol{\hat{\gamma}}_j;\boldsymbol{\hat{\gamma}}_j) \implies  f(\boldsymbol{\hat{\gamma}}_{j+1}) < f(\boldsymbol{\hat{\gamma}}_j), 
     \end{equation}
    Substituting~(\ref{eq:p_t_update}) in $g_i(\boldsymbol{\hat{\gamma}}_{j+1}[i];\boldsymbol{\hat{\gamma}}_j)$ gives us
     \begin{multline}
     \label{eq:g_i}
         g_i(\boldsymbol{\hat{\gamma}}_{j+1}[i];\boldsymbol{\hat{\gamma}}_j) = \boldsymbol{\hat{\gamma}}_j[i] ((\mathbf{T}_2(\boldsymbol{\hat{\gamma}}_j)[i])^{1-p}(\mathbf{T}_1(\boldsymbol{\hat{\gamma}}_j)[i])^{p} \\ + (\mathbf{T}_1(\boldsymbol{\hat{\gamma}}_j)[i])^{1-p}(\mathbf{T}_2(\boldsymbol{\hat{\gamma}}_j)[i])^{p})
     \end{multline}
     Substituting $r = \frac{1}{1-p}$ and $s = \frac{1}{p},$ we have $\frac{1}{r} + \frac{1}{s} = 1.$ We plan to use the generalized H\"{o}lders (GH) inequality~(Theorem 2 on page 19 of \cite{beckenbach2012inequalities}) which we state below for easy reference.\\
     {\it{Generalized H\"{o}lder Inequality:}} If $x_i, y_i \geq 0, s > 1, \frac{1}{s} + \frac{1}{r} =1,$ then 
     $$ 
     \sum_{i=1}^n x_i y_i \leq \left(\sum_{i=1}^n x_i^s\right)^{\frac{1}{s}} \left(\sum_{i=1}^n y_i^r\right)^{\frac{1}{r}}.
     $$
     The inequality is reversed if $s < 1, (s \neq 0).$ (For $s < 0,$ we assume that $x_i, y _i > 0.)$ In each case, the sign of equality holds if and only if the sets $(x^s)$ and $(y^r)$ are proportional.\\
     To use the GH inequality, we need $s > 1,$ which holds for $0<p<1$. Using the GH inequality, we obtain
     \begin{multline*}
         g_i(\boldsymbol{\hat{\gamma}}_{j+1}[i];\boldsymbol{\hat{\gamma}}_j) \leq \boldsymbol{\hat{\gamma}}_j[i]\left(\mathbf{T}_2(\boldsymbol{\hat{\gamma}}_j)[i] + \mathbf{T}_1(\boldsymbol{\hat{\gamma}}_j)[i]\right)^{\frac{1}{s}} \\ \times \left(\mathbf{T}_1(\boldsymbol{\hat{\gamma}}_j)[i] + \mathbf{T}_2(\boldsymbol{\hat{\gamma}}_j)[i]\right)^{\frac{1}{r}}=  \\\boldsymbol{\hat{\gamma}}_j[i]\left(\mathbf{T}_1(\boldsymbol{\hat{\gamma}}_j)[i] + \mathbf{T}_2(\boldsymbol{\hat{\gamma}}_j)[i]\right) =  g_i(\boldsymbol{\hat{\gamma}}_{j}[i];\boldsymbol{\hat{\gamma}}_j).
     \end{multline*}
     Taking the summation over $i$ on both sides gives us the condition in~\eqref{eq:mm_cond}. Since $\boldsymbol{\hat{\gamma}}_j$ is not a  stationary point, the condition for equality does not hold, and so we have strict descent for $0<p<1$.  So far the discussion does not include $p =1,$ and this choice requires special attention.
     
     Substituting $p=1$ in~\eqref{eq:g_i} gives  $g(\boldsymbol{\hat{\gamma}}_{j+1};\boldsymbol{\hat{\gamma}}_j)=g(\boldsymbol{\hat{\gamma}}_{j};\boldsymbol{\hat{\gamma}}_j)$. This suggests there is no increase in the negative log likelihood but does not assure descent. Note that the hyperplane in~\eqref{eq:logdet_mm} is a strict majorizer of the log-det term because the function $h(\boldsymbol{\gamma})$ defined as
     \begin{equation}
          h(\boldsymbol{\gamma};\boldsymbol{\hat{\gamma}}_j) = \sum^{M}_{i=1}\boldsymbol{\gamma}[i] \mathbf{T}_2(\boldsymbol{\hat{\gamma}}_j)[i] + c - \log(|\mm \mathbf{\Gamma} \mm^{H} + \sigma^2\mathbf{I}|),
     \end{equation}
     has the following Hessian
     \begin{equation}
         \nabla^2h(\boldsymbol{\gamma})= \left(\boldsymbol{\Phi}^{H}\Sigma^{-1}_{yy}\boldsymbol{\Phi}\right) \odot \left(\boldsymbol{\Phi}^{H}\Sigma^{-1}_{yy}\boldsymbol{\Phi}\right)^{*},
     \end{equation}
     where $\odot$ is the Hadamard product between two matrices. Note that the Hessian over $\mathbb{R}^{M}$ is positive semi-definite since the rank of $\boldsymbol{\Phi}$ is $N<<M$. But, the Hessian is positive definite over the domain $\boldsymbol{\gamma}\in \mathbb{R}^{M}_{+}$, because each element in the Hessian is strictly positive, therefore $\boldsymbol{\gamma}^T \nabla^2h(\boldsymbol{\gamma}) \boldsymbol{\gamma}>0, \forall \boldsymbol{\gamma}\neq 0$, since $\boldsymbol{\gamma}^T \nabla^2h(\boldsymbol{\gamma}) \boldsymbol{\gamma}$ is a sum of strictly positive elements. Hence, $h(\boldsymbol{\gamma})$ has a unique minimum of 0 and is attained only at $\boldsymbol{\gamma}=\boldsymbol{\hat{\gamma}}_j$. This implies that $g(\boldsymbol{\gamma};\boldsymbol{\hat{\gamma}}_j)$ is a strict majorizer of $f(\boldsymbol{\gamma})$. Even though $g(\boldsymbol{\hat{\gamma}}_{j+1};\boldsymbol{\hat{\gamma}}_j)=g(\boldsymbol{\hat{\gamma}}_{j};\boldsymbol{\hat{\gamma}}_j)$, because $g(.)$ is a strict majorizer we have that $f(\boldsymbol{\hat{\gamma}}_{j+1})<g(\boldsymbol{\hat{\gamma}}_{j+1})=g(\boldsymbol{\hat{\gamma}}_{j})=f(\boldsymbol{\hat{\gamma}}_{j})$. Hence, the case $p=1$ is also a valid update rule. 
     \subsection{Proof of Validity of EM Update Rule in $p$-SBL Majorizer}
\label{app:em_same_as_p}
This requires showing descent on the $p$-SBL majorizer, i.e. $\Delta_{j+1} = g_\text{$p$-SBL}(\boldsymbol{\hat{\gamma}}_{\text{EM},j+1};\boldsymbol{\hat{\gamma}}_j) - g_\text{$p$-SBL}(\boldsymbol{\hat{\gamma}}_j;\boldsymbol{\hat{\gamma}}_j) \leq 0$
Substituting the EM update rule into the majorizer corresponding to $p$-SBL and computing the difference results in (after some manipulations)
\begin{multline}
    \Delta_{j+1} = \sum^{M}_{i=1} \frac{\left( \mathbf{T}_1(\boldsymbol{\hat{\gamma}}_j)[i] - \mathbf{T}_2(\boldsymbol{\hat{\gamma}}_j)[i] \right)^2 \boldsymbol{\hat{\gamma}}^2_j[i]}{1+\left( \mathbf{T}_1(\boldsymbol{\hat{\gamma}}_j)[i] - \mathbf{T}_2(\boldsymbol{\hat{\gamma}}_j)[i] \right) \boldsymbol{\hat{\gamma}}_j[i]} \\ \times (\mathbf{T}_2(\boldsymbol{\hat{\gamma}}_j)[i]\boldsymbol{\hat{\gamma}}_j[i]-1). 
\end{multline}
Note that we can simplify the $i^{\text{th}}$ term in the summation, call it $\Delta^{(i)}_{j+1}$, by dividing and multiplying by $\boldsymbol{\hat{\gamma}}_j[i]$ to get the following 
\begin{equation}
    \Delta^{(i)}_{j+1} = \frac{\left( \mathbf{T}_1(\boldsymbol{\hat{\gamma}}_j)[i] - \mathbf{T}_2(\boldsymbol{\hat{\gamma}}_j)[i] \right)^2 \boldsymbol{\hat{\gamma}}^3_j[i]}{\boldsymbol{\hat{\gamma}}_{\text{EM},j+1}} (\mathbf{T}_2(\boldsymbol{\hat{\gamma}}_j)[i]\boldsymbol{\hat{\gamma}}_j[i]-1). 
\end{equation}
Clearly, the term multiplying $(\mathbf{T}_2(\boldsymbol{\hat{\gamma}}_j)[i]\boldsymbol{\hat{\gamma}}_j[i]-1)$ is non-negative. For the second term, we utilize the beamforming interpretation of $\mathbf{T}_2(\boldsymbol{\hat{\gamma}})$ in~\cite{mpdr_sbl} which shows that $1/\mathbf{T}_2(\boldsymbol{\hat{\gamma}})[i] - \boldsymbol{\hat{\gamma}}[i]$ represents the power of the residual term in each iteration of the MPDR (EM)-SBL algorithm. From this we can see that 
\begin{multline}
    1/\mathbf{T}_2(\boldsymbol{\hat{\gamma}})[i] - \boldsymbol{\hat{\gamma}}[i]\geq 0 \implies \mathbf{T}_2(\boldsymbol{\hat{\gamma}})[i]\boldsymbol{\hat{\gamma}}[i]\leq 1  \\ \implies \Delta^{(i)}_{j+1} \leq 0 \implies  \Delta_{j+1} \leq 0. 
\end{multline}


\ifCLASSOPTIONcaptionsoff
  \newpage
\fi



\bibliographystyle{IEEEtran}
\bibliography{bibtex/bib/IEEEexample}
%

%





\begin{IEEEbiography}[{\includegraphics[width=1in,height=1.25in,clip,keepaspectratio]{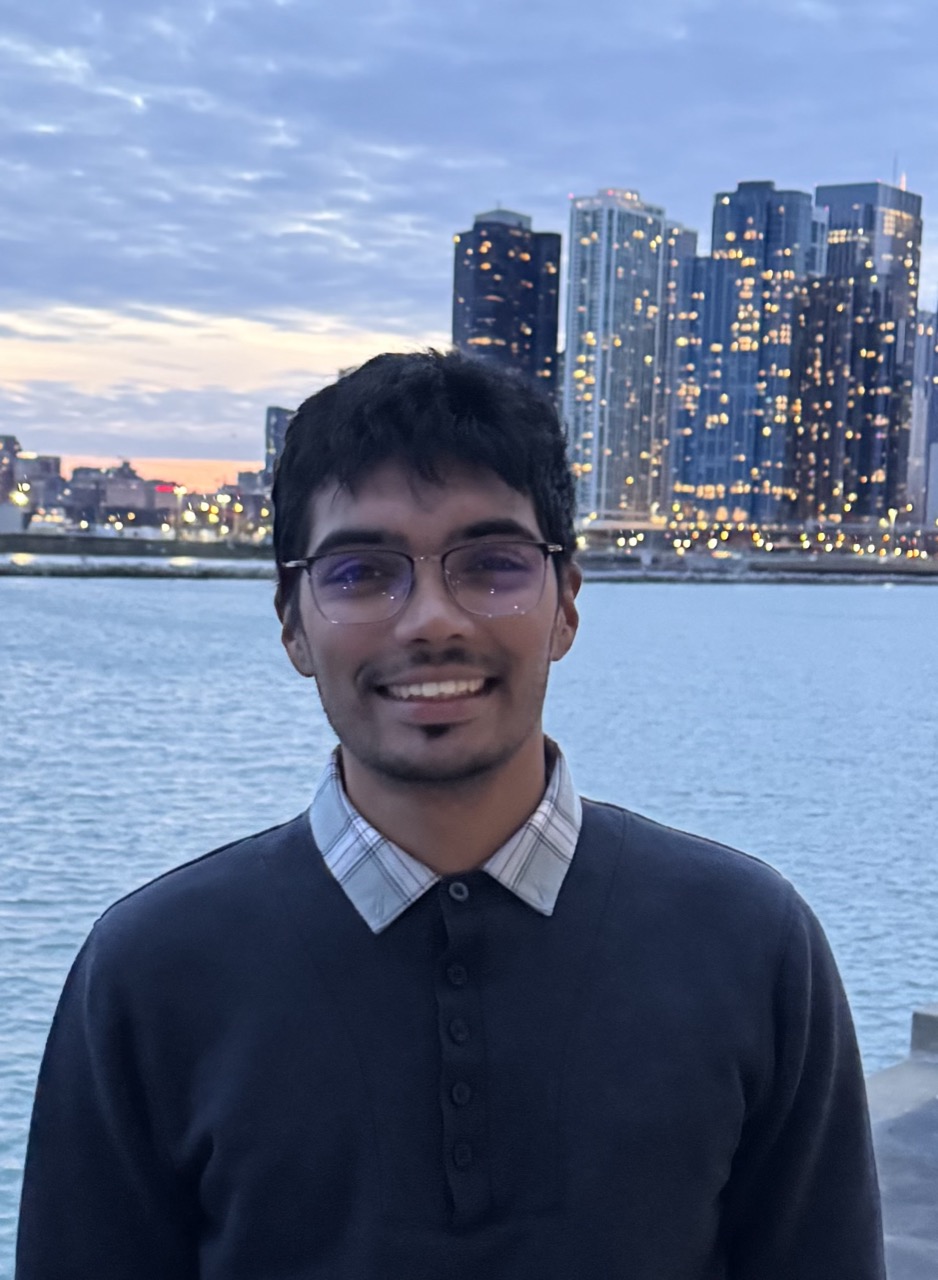}}]{Rushabha Balaji}
received his B.Eng in Electrical and Electronics Engineering from the Birla Institute of Technology and Science (BITS) Pilani, Pilani, in 2021, M.S. from University of California, San Diego (UCSD), La Jolla, CA, in 2024 and is currently pursuing his Ph.D from the University of California, Los Angeles (UCLA), Los Angeles, CA. He is the recipient of the UCLA graduate fellowship award in 2024 and the Kavli LIGO SURF award in 2021. His research interests span wireless communication, machine learning, non-convex optimization and array signal processing.
\end{IEEEbiography}

\begin{IEEEbiography}[{\includegraphics[width=1in,height=1.25in,clip,keepaspectratio]{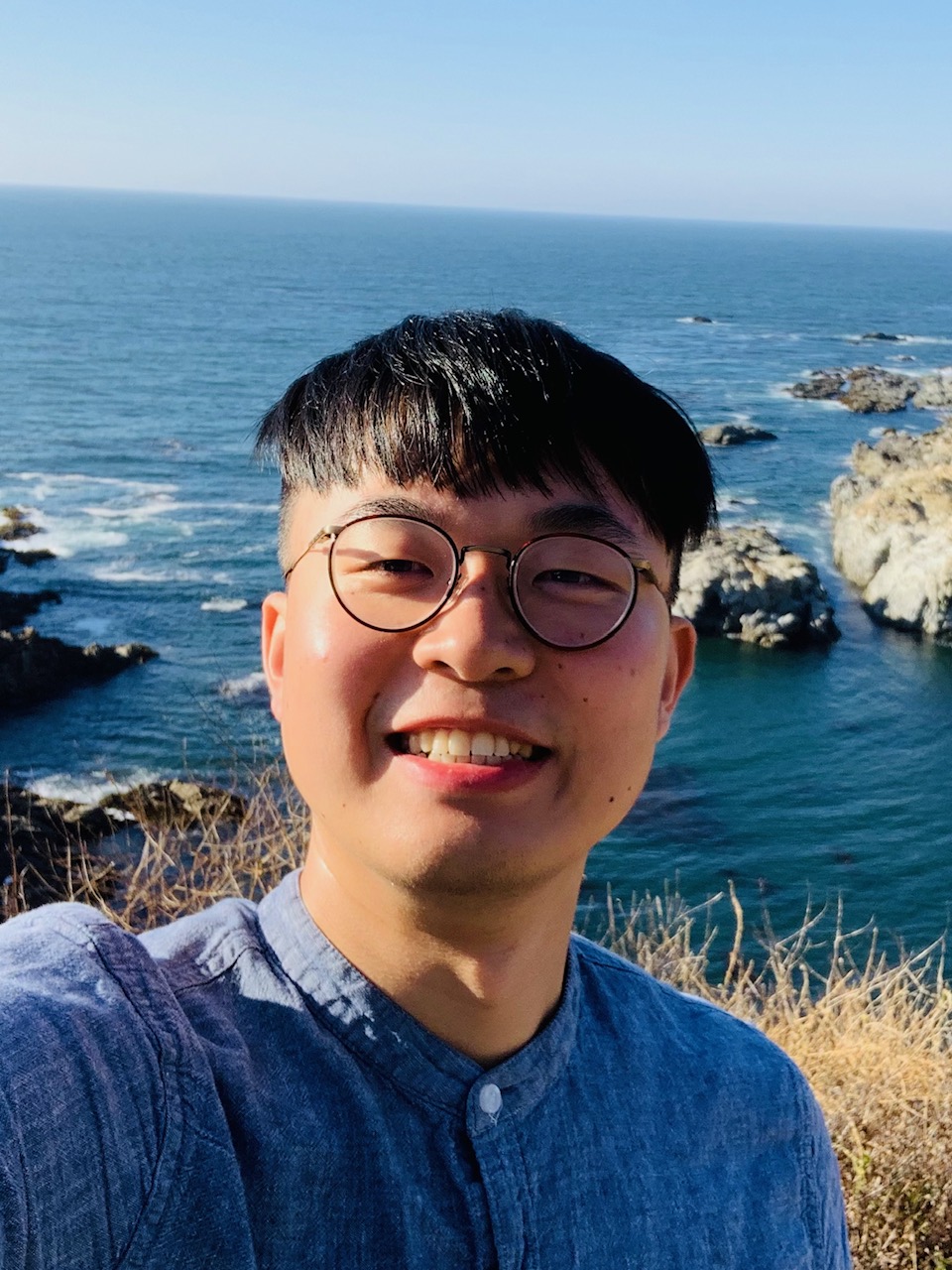}}]{Kuan-Lin Chen}
  (Member, IEEE) received the B.S. degree in electrical engineering from the National Taiwan University, Taipei, Taiwan, in 2016, and the M.S. and Ph.D. degrees from the University of California, San Diego (UCSD), La Jolla, CA, USA, in 2019 and 2024, respectively. After his graduate studies, he joined Apple as a Machine Learning Research Engineer. He was a recipient of the Qualcomm Innovation Fellowship, in 2022, the Innovative Research Grants Award from the Kavli Institute for Brain and Mind, in 2021, the IEEE Signal Processing Society Scholarship, in 2023, the NeurIPS Scholar Award, in 2022, the ICASSP Rising Stars in Signal Processing, in 2023, and the Best TA Award from the Department of Electrical and Computer Engineering, UCSD, in 2024. His research interests span several areas of machine learning and signal processing, including neural networks, deep learning, generative models, array processing, and speech processing.
\end{IEEEbiography}

\begin{IEEEbiography}[{\includegraphics[width=1in,height=1.25in,clip,keepaspectratio]{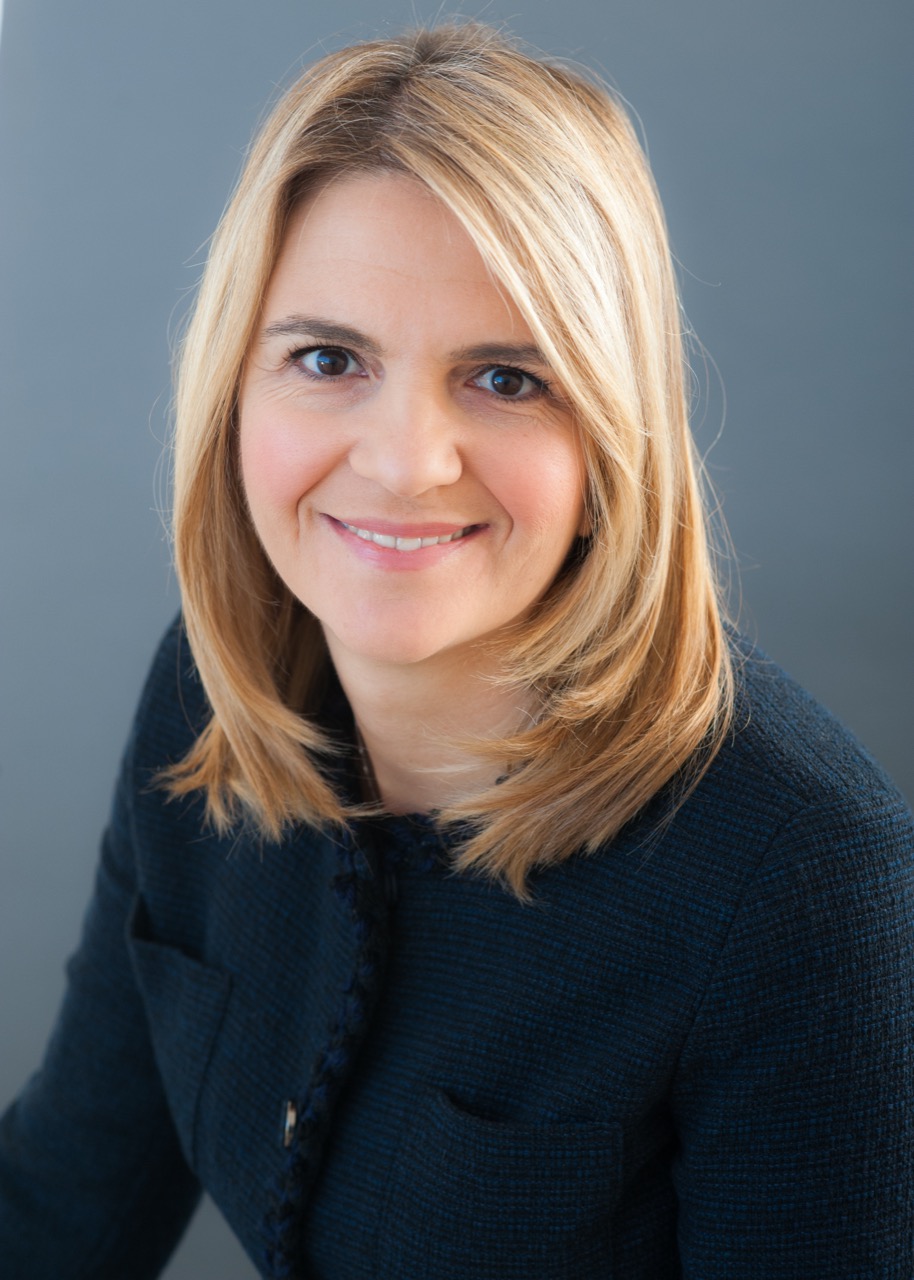}}]{Danijela Cabric}
is a Professor in the Electrical and Computer Engineering Department at the University of California, Los Angeles. She received M.S. from the University of California, Los Angeles in 2001 and Ph.D. from University of California, Berkeley in 2007, both in Electrical Engineering. In 2008, she joined UCLA as an Assistant Professor, where she heads Cognitive Reconfigurable Embedded Systems lab. Her current research projects include novel radio architectures, signal processing, communications, machine learning and networking techniques for spectrum sharing, 5G millimeter-wave, massive MIMO and IoT systems. Prof. Cabric was a recipient of the Samueli Fellowship in 2008, the Okawa Foundation Research Grant in 2009, Hellman Fellowship in 2012 and the National Science Foundation Faculty Early Career Development (CAREER) Award in 2012, and the Qualcomm Faculty Award in 2020 and 2021. Prof. Cabric is an IEEE Fellow.
\end{IEEEbiography}

\begin{IEEEbiography}[{\includegraphics[width=1in,height=1.25in,clip,keepaspectratio]{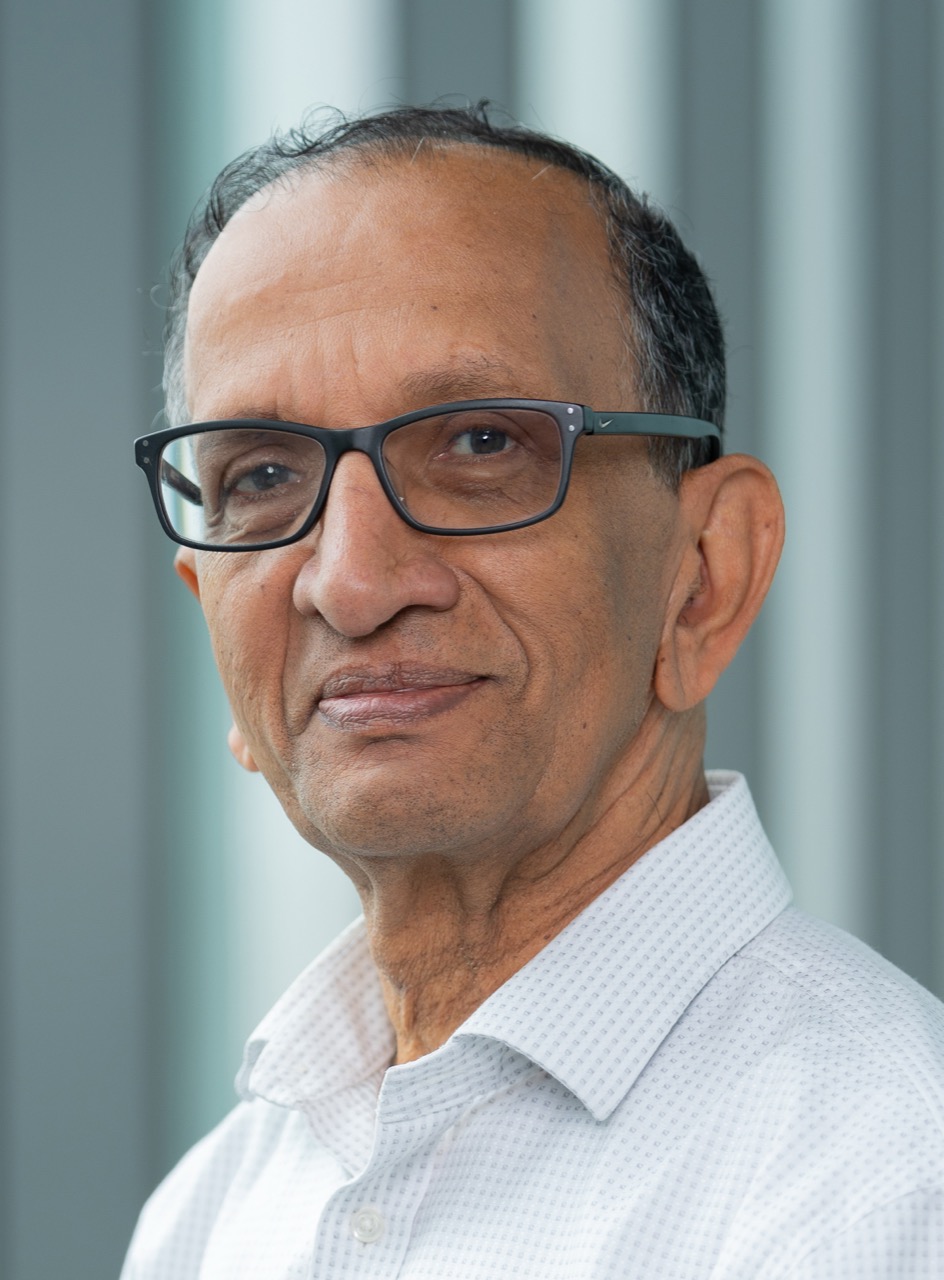}}]{Bhaskar D. Rao}
(Life Fellow, IEEE) received his B.Tech. degree in Electronics and Electrical Communication Engineering from the Indian Institute of Technology, Kharagpur, India, in 1979 and his M.S. and Ph.D. degrees from the University of Southern California, Los Angeles in 1981 and 1983, respectively. He has been teaching and conducting research at the University of California in San Diego, La Jolla since 1983, where he is currently a Professor Emeritus and Distinguished Professor of the Graduate Division in the Electrical and Computer Engineering department. He has also been the holder of the Ericsson Endowed Chair in Wireless Access Networks and Distinguished Professor and the Director of the Center for Wireless Communications (2008-2011).

Professor Rao's research interests are in the areas of statistical signal processing, estimation theory, optimization theory and machine learning, with applications to digital communications, speech signal processing, and biomedical signal processing. He is a pioneer in the theory and use of sparsity in signal processing applications. Since co-authoring the first paper on the seminal FOCUSS algorithm in 1992, he has been driving the field of sparsity forward, including co-organizing the first special session on sparsity at ICASSP 1998 entitled "SPEC-DSP: Signal Processing with Sparseness Constraint." His work has received several paper awards, including the 2012 Signal Processing Society (SPS) best paper award for the paper "An Empirical Bayesian Strategy for Solving the Simultaneous Sparse Approximation Problem," with David P. Wipf and the Stephen O. Rice Prize paper award in the field of communication systems for the paper "Network Duality for Multiuser MIMO Beamforming Networks and Applications," with B. Song and R. L. Cruz. Professor Rao was elected fellow of IEEE in 2000 for his contributions to the statistical analysis of subspace algorithms for harmonic retrieval. He received the IEEE Signal Processing Society Technical Achievement Award in 2016 and was the recipient of the 2023 IEEE SPS Norbert Wiener Society Award. He has been a member of the Statistical Signal and Array Processing Technical Committee, the Signal Processing Theory and Methods Technical Committee, the Communications Technical Committee of the IEEE Signal Processing Society, SPS Fellow Evaluation Committee (2023-2024) and was the chair of the Machine Learning for Signal Processing Technical Committee (2019-2020).
\end{IEEEbiography}

\end{document}